\documentclass[aps,prd,groupedaddress,nofootinbib]{revtex4-2}

\pdfoutput=1

\usepackage{float}
\usepackage{amsmath}
\usepackage{color}
\usepackage{graphicx}
\usepackage[dvipsnames]{xcolor}
\usepackage{url}
\usepackage{epsfig}
\usepackage[T1]{fontenc}
\usepackage{multirow}
\usepackage{physics} 
\usepackage{booktabs} 
\usepackage{array} 
\usepackage{paralist} 
\usepackage{grffile}
\usepackage{verbatim} 
\usepackage{subfig} 
\usepackage{amsmath,amsthm,amssymb,bm,amsfonts}
\usepackage{slashed}
\usepackage[utf8]{inputenc}
\usepackage{hyperref}
\allowdisplaybreaks[1]

\usepackage{color}
\usepackage[normalem]{ulem}

\usepackage{multirow}
\usepackage[title]{appendix}

\def\beg{\begin{equation}}
\def\eeg{\end{equation}}
\def\bea{\begin{eqnarray}}
\def\eea{\end{eqnarray}}

\newcommand{\slv}{\raise.15ex\hbox{$/$}\kern-.53em\hbox{$v$}}
\newcommand{\slnbar}{\raise.15ex\hbox{$/$}\kern-.53em\hbox{$\bar{n}$}}
\newcommand{\slF}{\raise.15ex\hbox{$/$}\kern-.53em\hbox{$F$}}
\newcommand{\sllbar}{\raise.15ex\hbox{$/$}\kern-.40em\hbox{$\bar{l}$}}
\newcommand{\slh}{\raise.15ex\hbox{$/$}\kern-.40em\hbox{$h$}}
\newcommand{\slP}{\raise.15ex\hbox{$/$}\kern-.53em\hbox{$P$}}
\newcommand{\slR}{\raise.15ex\hbox{$/$}\kern-.53em\hbox{$R$}}
\newcommand{\slz}{\raise.15ex\hbox{$/$}\kern-.53em\hbox{$Z$}}
\newcommand{\slzbar}{\raise.15ex\hbox{$/$}\kern-.53em\hbox{$\bar{Z}$}}
\newcommand{\slQ}{\raise.15ex\hbox{$/$}\kern-.53em\hbox{$Q$}}
\newcommand{\slK}{\raise.15ex\hbox{$/$}\kern-.53em\hbox{$K$}}
\newcommand{\slkbar}{\raise.15ex\hbox{$/$}\kern-.53em\hbox{$\bar{k}$}}
\newcommand{\slkone}{\raise.15ex\hbox{$/$}\kern-.53em\hbox{$k_1$}}
\newcommand{\slpone}{\raise.15ex\hbox{$/$}\kern-.53em\hbox{$p_1$}}
\newcommand{\slpbarone}{\raise.15ex\hbox{$/$}\kern-.53em\hbox{$\bar{p}_1$}}
\newcommand{\slptwo}{\raise.15ex\hbox{$/$}\kern-.53em\hbox{$p_2$}}
\newcommand{\slpbartwo}{\raise.15ex\hbox{$/$}\kern-.53em\hbox{$\bar{p}_2$}}
\newcommand{\slqone}{\raise.15ex\hbox{$/$}\kern-.53em\hbox{$q_1$}}
\newcommand{\slD}{\raise.15ex\hbox{$/$}\kern-.53em\hbox{$\!D$}}
\newcommand{\slC}{\raise.15ex\hbox{$/$}\kern-.53em\hbox{$C$}}
\newcommand{\slA}{\raise.15ex\hbox{$/$}\kern-.73em\hbox{$A$}}
\newcommand{\slSigma}{\raise.15ex\hbox{$/$}\kern-.53em\hbox{$\Sigma$}}
\newcommand{\slpartial}{\raise.15ex\hbox{$/$}\kern-.53em\hbox{$\partial$}}
\newcommand{\slcalP}{\raise.15ex\hbox{$/$}\kern-.63em\hbox{$\cal P$}}
\newcommand{\sleps}{\raise.15ex\hbox{$/$}\kern-.53em\hbox{$\epsilon$}}
\newcommand{\slepsbar}{\raise.15ex\hbox{$/$}\kern-.53em\hbox{$\overline{\epsilon}$}}
\newcommand{\slepsstar}{\raise.15ex\hbox{$/$}\kern-.53em\hbox{$\epsilon$}^\star}
\newcommand{\slS}{\raise.15ex\hbox{$/$}\kern-.73em\hbox{$S$}}

\newcommand{\eps}{\varepsilon}
\newcommand{\bb}{\mathbf}

\newcommand{\bk}{\mathbf{k}}
\newcommand{\bp}{\mathbf{p}}
\newcommand{\bq}{\mathbf{q}}
\newcommand{\bx}{\mathbf{x}}

\newcommand{\dtwo}[1]{\frac{\dd^2 #1}{(2\pi)^2}}

\newcommand{\p}{\prime}

\begin{document}
\title{SIDIS at small $x$ at next-to-leading order: Gluon contribution}

\author{Filip Bergabo$^{a}$, Jamal Jalilian-Marian$^{a,b}$}
\affiliation{$^{a}$Department of Natural Sciences, Baruch College, CUNY, 17 Lexington Avenue, New York, New York 10010, USA\\
$^{b}$City University of New York Graduate Center, 365 Fifth Avenue, New York, New York  10016, USA}

\begin{abstract}
We calculate the contribution of gluons to single inclusive hadron production at next-to-leading order (NLO) accuracy in deep inelastic scattering (DIS) at small $x$ using the color glass condensate formalism. It is shown that the only divergence present is the standard collinear divergence, which is absorbed into scale evolution of quark-hadron fragmentation function. Our calculations are performed at finite $N_c$ and we provide general finite $N_c$ expressions for the structure of Wilson lines appearing in inclusive dihadron and single hadron production cross sections. We also comment on how one can obtain rapidity distribution of hadron multiplicities from our results.
\end{abstract}

\maketitle



\section{Introduction}
Deep inelastic scattering (DIS) is the cleanest environment in which to probe
internal partonic structure of hadrons and nuclei~\cite{Aschenauer:2016our,Accardi:2012qut}. 
Due to the experimentally 
observed fast rise of gluons at small Bjorken $x$ one expects that a high 
energy hadron or nucleus is a maximum occupancy state of predominantly 
gluons~\cite{Gribov:1983ivg,Mueller:1985wy}. 
This high density gluonic state is referred to as the color glass condensate 
(CGC)~\cite{McLerran:1993ni,McLerran:1993ka,Jalilian-Marian:1996mkd}
and is expected to give the dominant contribution to scattering cross sections at high
energy. While there have been tantalizing hints of this saturated state
at HERA, RHIC and the 
LHC~\cite{Jalilian-Marian:2004vhw,Dumitru:2005gt,Jalilian-Marian:2005qbq,Marquet:2007vb,Albacete:2010pg,Stasto:2011ru,Lappi:2012nh,Jalilian-Marian:2012wwi,Jalilian-Marian:2011tvq,Zheng:2014vka,Stasto:2018rci,Albacete:2018ruq,Mantysaari:2019hkq,Hatta:2020bgy,Jia:2019qbl,Gelis:2002fw,Dominguez:2011wm,Metz:2011wb,Dominguez:2011br,Iancu:2013dta,Altinoluk:2015dpi,Hatta:2016dxp,Dumitru:2015gaa,Kotko:2015ura,Marquet:2016cgx,vanHameren:2016ftb,Marquet:2017xwy,Dumitru:2018kuw,Dumitru:2001jn,Dumitru:2002qt,Mantysaari:2019csc,Salazar:2019ncp,Boussarie:2021ybe,Ayala:1995hx,Jalilian-Marian:2004cdc,Kotko:2017oxg,Hagiwara:2017fye,Henley:2005ms,Klein:2019qfb,Hatta:2021jcd,Kolbe:2020tlq,Gelis:2002fw,Jalilian-Marian:2005tod,Altinoluk:2019fui,Boussarie:2019ero,Boussarie:2016ogo,Boussarie:2014lxa,Dumitru:2010ak} its presence is far from being firmly established 
(see \cite{Iancu:2003xm,Jalilian-Marian:2005ccm,Weigert:2005us,Gelis:2010nm,Morreale:2021pnn} for reviews). 
The
proposed Electron Ion Collider (EIC) will enable us to look for gluon saturation in
various channels with unprecedented precision and accuracy. Among the most
promising processes in which gluon saturation is expected to play a dominant 
role are single and double inclusive jet and hadron production. Measurements 
of transverse momenta, angular correlations and rapidity dependence of these processes
will shed light on dynamics of gluons saturation and its energy dependence and enable a 
detailed comparison with theoretical expectations based on CGC formalism~\cite{Balitsky:1995ub,Kovchegov:1999yj,Jalilian-Marian:1997qno,Jalilian-Marian:1997jhx,Jalilian-Marian:1997ubg,Jalilian-Marian:1998tzv,Kovner:1999bj,Kovner:2000pt,Iancu:2000hn,Ferreiro:2001qy}. 
To improve the
theoretical accuracy of CGC-based predictions there has been significant progress 
made in calculating next to leading order (NLO) and beyond eikonal corrections to  observable which are sensitive to gluon saturation~\cite{Fadin:1998py,Chirilli:2011km,Chirilli:2012jd,balitsky:2012bs,Balitsky:2013fea,grabovsky:2013mba,caron-huot:2013fea,kovner:2013ona,lublinsky:2016meo,Caucal:2022ulg,Caucal:2023nci,Caucal:2023fsf,caron-huot:2016tzz,Fucilla:2023mkl,Fucilla:2022wcg,boussarie:2017dmx,beuf:2022ndu,beuf:2021srj,beuf:2021qqa,mantysaari:2021ryb,mantysaari:2022bsp,mantysaari:2022kdm,lappi:2021oag,Ayala:2016lhd,Ayala:2017rmh,Ayala:2014nza,Iancu:2020mos,roy:2019hwr,hatta:2022lzj,Iancu:2021rup,Iancu:2020jch,Taels:2022tza,Caucal:2021ent,Bergabo:2021woe,Bergabo:2022tcu,Kovner:2001vi,Kovchegov:2017lsr,Cougoulic:2019aja,Kovchegov:2018znm,Kovchegov:2017jxc,Kovchegov:2016zex,Kovchegov:2016weo,Kovchegov:2015pbl,Agostini:2019hkj,Agostini:2019avp,Altinoluk:2015xuy,Altinoluk:2015gia,Altinoluk:2014oxa,jalilian-marian:2021lhe,Jalilian-Marian:2019kaf,Jalilian-Marian:2018iui,Jalilian-Marian:2017ttv,Hentschinski:2017ayz,Hentschinski:2016wya,Gituliar:2015agu,Balitsky:2016dgz,Balitsky:2015qba,Mukherjee:2023snp,Fu:2023jqv,Boussarie:2021wkn,Boussarie:2023xun}. 

In~\cite{Bergabo:2022zhe} we calculated the next-to-leading order corrections to single 
inclusive hadron production in DIS (SIDIS) at small $x$ when either quark or antiquark 
in the final state hadronizes. Here we continue our work toward a complete NLO 
calculation of SIDIS by considering hadronization of the produced gluon~\footnote{While this manuscript was being prepared for publication we were informed of a similar work~\cite{Caucal:2024cdq} which however focuses on single inclusive jet production in DIS. We thank F. Salazar for bringing their work to our attention.}. This channel is 
not present in a leading order calculations of SIDIS and appears only as part of the NLO
corrections so that it is important to understand its role and contribution. Furthermore
we include the full $N_c$ dependence rather than making the large $N_c$ approximation.

In the small $x$ limit of DIS the virtual photon (transverse or longitudinal) splits into a quark antiquark pair (a dipole), which then multiply scatters from the target hadron or nucleus. To leading order (LO)  accuracy the double inclusive production cross section can be written as  

\bea
\frac{\dd \sigma^{\gamma^* p/A \to q\bar{q} X}}
{\dd^2 \bb{p}\, \dd^2 \bb{q} \, \dd y_1 \, \dd y_2} &=& 
\frac{ e^2 Q^2 N_c}{(2\pi)^7} \delta(1-z_1-z_2) \, (z_1 z_2)^2 \, 
\int \dd^8 \bx \left[S_{122^\prime 1^\prime} - S_{12} - S_{1^\prime 2^\prime} + 1\right]
\,
e^{i\bb{p}\cdot\bb{x}_{1^\p1}} 
e^{i\bb{q}\cdot\bb{x}_{2^\p2}} 
\nonumber \\
&&  
\bigg\{4z_1z_2K_0(|\bb{x}_{12}|Q_1) 
K_0(|\bb{x}_{1^\prime 2^\prime}|Q_1) + 
(z_1^2 + z_2^2) \,
\frac{ \bb{x}_{12}\cdot \bb{x}_{1^\prime 2^\prime}}{|\bb{x}_{12}| |\bb{x}_{1^\prime 2^\prime}|} \, 
K_1(|\bb{x}_{12}|Q_1)K_1(|\bb{x}_{1^\prime 2^\prime}|Q_1) 
\bigg\} \label{LOdsig}
\eea
where the first and second terms in the curly bracket above correspond to the contribution of the longitudinal and transverse polarizations of the virtual photon. 
The incoming virtual photon (right moving) has momentum $l^\mu$ with $l^2 = -\, Q^2$ and we have set the 
transverse momentum of the photon to zero without any loss of generality.
Furthermore $p^\mu$ ($q^\mu$) is the momentum of the outgoing quark (antiquark) while $z_1$ ($z_2$) is its longitudinal momentum fraction relative to the photon. Also, 
$\bx_1$ ($\bx_2$) is the transverse coordinate of the quark (antiquark), and primed coordinates are used in the conjugate amplitude. Quark and antiquark rapidities $y_1$ and $y_2$ are related to their momentum fractions $z_1$ and $z_2$ via $\dd y_i = \dd z_i / z_i$. 

We further define and use the notation 
\begin{align}
Q_i = Q\sqrt{z_i(1-z_i)}, \,\,\,\,\,\, \bx_{ij} = \bx_i - \bx_j,\,\,\,\,\,\, \dd^8 \bx = \dd^2 \bx_1 \, \dd^2 \bx_2\, \dd^2 \bx_{1^\p} \, \dd^2 \bx_{2^\p}.
\end{align}
The production cross section is a convolution of the probability for a photon to split into a quark at transverse position $\bx_1$ and an antiquark at position $\bx_2$ represented by the Bessel functions, with multiple scatterings of the quark antiquark 
pair on the target encoded in the dipoles $S_{ij}$ and quadrupoles $S_{ijkl}$. 
To calculate the single inclusive production cross section we need to integrate over one of the final state partons which we choose to be antiquark so that quark is produced. The case when antiquark is produced is identical so that our results can be multiplied by a factor of $2$ to include antiquark production. Integrating over the produced antiquark phase space sets $\bx_{2^\p} = \bx_2$ and gives 

\bea
\frac{\dd \sigma^{\gamma^* p/A \to q (\bp,y_1)X}}{\dd^2 \bp\, \dd y_1} &=&
\frac{ e^2 Q^2 N_c}{(2\pi)^5} \int \dd z_2 \, \delta (1-z_1-z_2)\, (z_1^2 \, z_2)\, 
\int \dd^6 \bx \left[S_{11^\p} - S_{12} - S_{1^\p2} + 1\right] 
e^{i\bp\cdot\bx_{1^\p1}} 
\nonumber \\
&&
\bigg\{
4z_1z_2 K_0(|\bx_{12}|Q_1) K_0(|\bx_{1^\p2}|Q_1) 
+ 
(z_1^2 + z_2^2) \,
\frac{\bx_{12}\cdot \bx_{1^\p2}} {|\bx_{12}||\bx_{1^\p2}|} \, 
K_1(|\bx_{12}|Q_1) K_1(|\bx_{1^\p2}|Q_1) 
\bigg\} .
\label{LOdsig-sidis-quark}
\eea

\section{Single inclusive gluon production}
\begin{figure}[H]
\centering
\includegraphics[width=70mm]{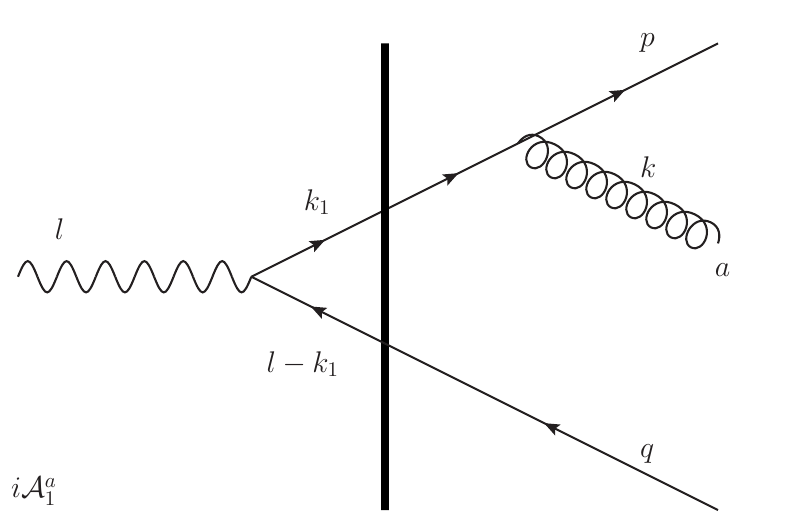}\includegraphics[width=70mm]{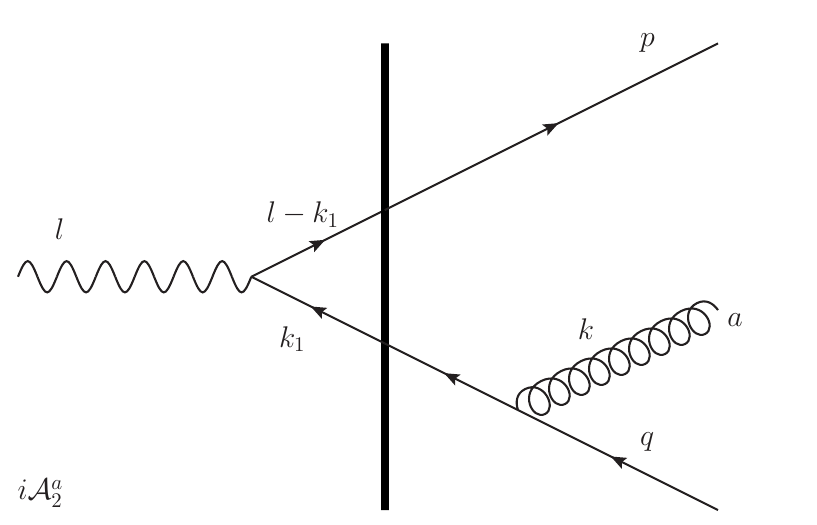}\\ \includegraphics[width=70mm]{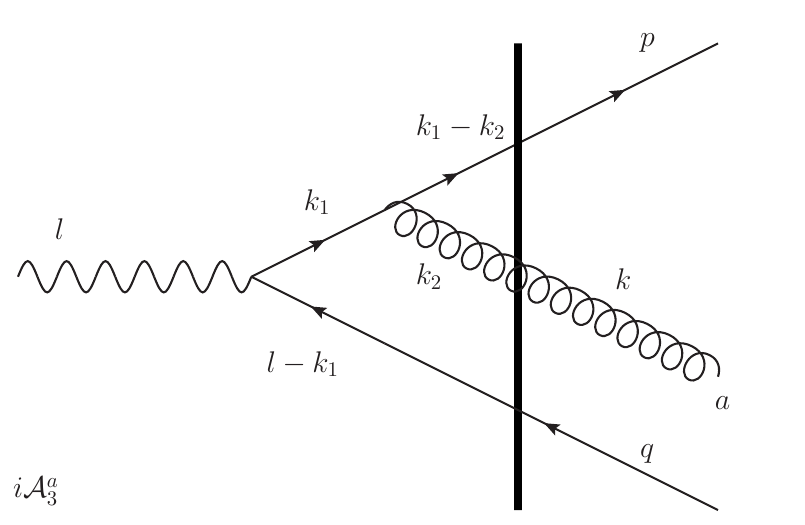}\includegraphics[width=70mm]{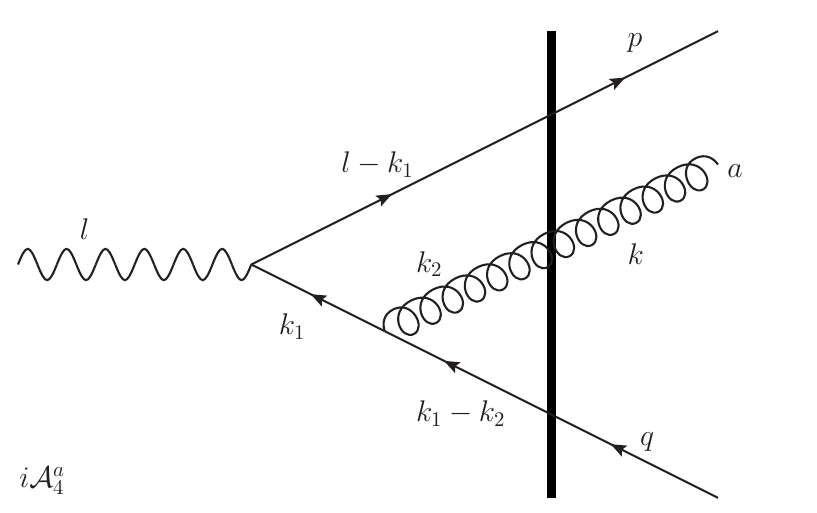}
\caption{The real corrections $i\mathcal{A}_1^a, ..., i\mathcal{A}_4^a$. The arrows on Fermion lines indicate Fermion number flow, whereas all momenta flow to the right. The thick solid line indicates multiple scatterings from the target.}\label{fig:realdiags}
\end{figure}

To calculate contribution of gluons to SIDIS we start with the next-to-leading (NLO) corrections to dihadron production in~\cite{Bergabo:2022tcu}. In principle there are 
both real and virtual contributions, however only real diagrams containing a gluon in the
final state contribute so therefore we will focus on the real corrections here. The diagrams corresponding to real corrections are shown in Fig.~\ref{fig:realdiags},

The full NLO expressions for dihadron production are derived in~\cite{Bergabo:2022tcu}, 
in order to calculate contribution of gluons one needs to "undo" integration of 
the final state gluon phase space and then integrate out the final state quark and 
antiquark so that only the produced gluon is left in the final 
state. As the original expressions in~\cite{Bergabo:2022tcu} are very long and not 
illuminating we will just show the results here,

\begin{align}
\frac{\dd \sigma^{L}_{1\times 1}}{\dd^2 \bk\, \dd y} = &
4 \frac{e^2 Q^2 g^2 N_c}{(2\pi)^{6}}\,  C_F 
\int \dd z_1 \dd z_2 \, \delta ( 1 - z_1 - z_2 - z) \, 
z_2^2 (1 - z_2)^2 \left[z_1^2+(1 - z_2)^2\right]
\nonumber \\ 
&
\int 
\dd^{6} \bx \, e^{i\bk\cdot\bx_{1^\p 1}} \,
K_0(|\bx_{12}|Q_2) K_0(|\bx_{1^\p 2}|Q_2) \, 
\left[ S_{1 1^\p} - S_{12} - S_{1^\p 2} + 1\right] 
\int \dtwo{\bp} \frac{e^{i\bp\cdot\bx_{1^\p 1}}}
{\left[z \bp - z_1 \bk\right]^2} 
\label{eq:onexone}
\\ 
\frac{\dd \sigma^{L}_{2\times 2}}{\dd^2 \bk\, \dd y} = & 
4 \frac{e^2 Q^2 g^2 N_c}{(2\pi)^{6}} \, C_F 
\int \dd z_1 \dd z_2  \, \delta ( 1 - z_1 - z_2 - z) \,  
z_1^2 (1 - z_1)^2 \left[z_2^2+(1 - z_1)^2\right] 
\nonumber \\ 
&
\int \dd^{6} \bx \, e^{i\bk\cdot\bx_{2^\p2}} \, 
K_0(|\bx_{12}|Q_1)K_0(|\bx_{12^\p}|Q_1) \, \left[ S_{2 2^\p} - S_{12} - S_{12^\p} + 1\right] 
\int \dtwo{\bq} \frac{e^{i\bq\cdot\bx_{2^\p2}}}
{\left[z \bq - z_2 \bk\right]^2} 
\label{eq:twoxtwo}
\\
\nonumber\\
\frac{\dd \sigma^{L}_{1\times 2}}{\dd^2 \bk\, \dd y} = & 
4 \frac{e^2 Q^2 g^2 N_c}{(2\pi)^{6}} \int \dd z_1 \dd z_2 \,
\delta ( 1 - z_1 - z_2 - z) \,
\left[\frac{z_1 (1-z_1) z_2 (1-z_2)}{z}\right]
\left[\frac{z_1 (1- z_1) + z_2 (1 - z_2)}{z}\right] \nonumber\\
&
\frac{1}{(2 \pi)^2}  
\int \dd^{8} \bx\, 
\frac{\bx_{1^\p 1}\cdot\bx_{2^\p 2}}{\bx_{1^\p 1}^2 \bx_{2^\p 2}^2}\,
K_0(|\bx_{12}|Q_2) \, K_0(|\bx_{1^\p2^\p}| Q_1) \,
e^{i \bk \cdot\left[\bx_{2^\p 1} + 
\frac{z_1}{z} \bx_{1^\p 1} + \frac{z_2}{z} \bx_{2^\p 2}\right]}
\nonumber \\
&
\left[\frac{N_c}{2} S_{12}S_{1^\p2^\p} + C_F\, [1 - S_{12} - S_{2^\p1^\p}]  
- \frac{1}{2 N_c} S_{122^\p1^\p}\right]
\label{eq:onextwo}
\\
\frac{\dd \sigma^{L}_{3\times 3}}{\dd^2 \bk\, \dd y} = & 
4 \frac{e^2 Q^2 g^2 N_c}{(2\pi)^{6}} \int \dd z_1 \dd z_2 \,
\delta ( 1 - z_1 - z_2 - z) \, z_2^2 
\left[z_1^2 + (1 - z_2)^2\right] \nonumber\\
&
\frac{1}{(2 \pi)^2}  
\int \dd^{8} \bx \, 
\frac{\bx_{31}\cdot\bx_{3^\p1}}{\bx_{31}^2 \bx_{3^\p1}^2}\,
\left[K_0(Q X) \, K_0(Q X^\p)\right]_{1^\p = 1 \, ,\, 2^\p = 2} \,
e^{i \bk \cdot\bx_{3^\p 3}}
 \nonumber \\
&
\left\{\frac{N_c}{2} \left[S_{33^\p} S_{33^\p} - S_{13} S_{23} - S_{13^\p} S_{23^\p}\right] 
+ C_F - \frac{1}{2 N_c} \left[1 - 2 S_{12}\right]\right\}.\label{Real33}\\
\frac{\dd \sigma^{L}_{4\times 4}}{\dd^2 \bk\, \dd y} = & 
4 \frac{e^2 Q^2 g^2 N_c}{(2\pi)^{6}} \int \dd z_1 \dd z_2 \, 
\delta ( 1 - z_1 - z_2 - z) \, 
z_1^2 \left[z_2^2 + (1 - z_1)^2\right] \nonumber\\
&
\frac{1}{(2 \pi)^2}  
\int \dd^{8} \bx \, 
\frac{\bx_{3 2}\cdot\bx_{3^\p 2}}{\bx_{3 2}^2 \bx_{3^\p 2}^2}\,
\left[K_0(Q X) \, K_0(Q X^\p)\right]_{1^\p = 1 \, , \, 2^\p = 2} \,
e^{i \bk \cdot\bx_{3^\p 3}}
\nonumber \\
&
\left\{\frac{N_c}{2} \left[S_{33^\p} S_{33^\p} - S_{13} S_{23} - S_{13^\p} S_{23^\p}\right] 
+ C_F - \frac{1}{2 N_c} \left[1 - 2\, S_{12}\right]\right\}.\label{Real44}\\
\frac{\dd \sigma^{L}_{3\times 4}}{\dd^2 \bk\, \dd y} = & 
- 4 \frac{e^2 Q^2 g^2 N_c}{(2\pi)^{6}} \int \dd z_1 \dd z_2\,  
\delta ( 1 - z_1 - z_2 - z) \, 
z_1 z_2 \left[z_1 (1 - z_1) + z_2 (1 - z_2) \right] 
\nonumber\\
&
\frac{1}{(2 \pi)^2}  
\int \dd^{8} \bx  \, 
\frac{\bx_{3 1}\cdot\bx_{3^\p 2}}{\bx_{3 1}^2 \bx_{3^\p 2}^2}\,
\left[K_0(Q X) \, K_0(Q X^\p)\right]_{1^\p = 1 \, , \, 2^\p = 2} \,
e^{i \bk \cdot\bx_{3^\p 3}}
 \nonumber \\
&
\left\{\frac{N_c}{2} \left[S_{33^\p} S_{33^\p} - S_{13} S_{23} - S_{13^\p} S_{23^\p}\right] 
+ C_F - \frac{1}{2 N_c} \left[1 - 2\, S_{12}\right]\right\}.\label{Real34}\\
\frac{\dd \sigma^{L}_{1\times 3}}{\dd^2 \bk\, \dd y} = & 
4 \frac{e^2 Q^2 g^2 N_c}{(2\pi)^{6}} \int \dd z_1 \dd z_2 \,
\delta ( 1 - z_1 - z_2 - z) \,
\frac{z_2}{z} z_2 (1- z_2) \left[z_1^2 + (1 - z_2)^2\right] 
\nonumber\\
&
\frac{1}{(2 \pi)^2}  
\int \dd^{8} \bx \, 
\frac{\bx_{1^\p1}\cdot\bx_{3^\p 1^\p}}{\bx_{1^\p1}^2 \bx_{3^\p 1^\p}^2}\,
\left[K_0(|\bx_{12}|Q_2) \, K_0(Q X^\p)\right]_{2^\p = 2} \,
e^{i \bk \cdot\left[\bx_{3^\p 1} +\frac{z_1}{z} \bx_{1^\p1}\right]}
\nonumber \\
&
\left\{\frac{N_c}{2} \left[S_{3^\p1^\p} S_{13^\p} - S_{3^\p1^\p} S_{23^\p}\right] 
+ C_F \left[1 - S_{12}\right] - \frac{1}{2 N_c} \left[S_{11^\p} - S_{21^\p}\right]\right\}.\label{Real13}\\
\frac{\dd \sigma^{L}_{1\times 4}}{\dd^2 \bk\, \dd y} = & 
- 4 \frac{e^2 Q^2 g^2 N_c}{(2\pi)^{6}} \int \dd z_1 \dd z_2 \,
\delta ( 1 - z_1 - z_2 - z) \,
\frac{z_1}{z} z_2 (1- z_2) 
\left[z_1 (1 - z_1) + z_2 (1 - z_2)\right] \nonumber\\
&
\frac{1}{(2 \pi)^2}  
\int \dd^{8} \bx \, 
\frac{\bx_{1^\p1}\cdot\bx_{3^\p 2}}{\bx_{1^\p1}^2 \bx_{3^\p 2}^2}\,
\left[K_0(|\bx_{12}|Q_2) \, K_0(Q X^\p)\right]_{2^\p = 2} \,
e^{i \bk \cdot\left[\bx_{3^\p 1} +\frac{z_1}{z} \bx_{1^\p1}\right]}
\nonumber \\
&
\left\{\frac{N_c}{2} \left[S_{3^\p1^\p} S_{13^\p} - S_{3^\p1^\p} S_{23^\p}\right] 
+ C_F \left[1 - S_{12}\right] - \frac{1}{2 N_c} \left[S_{11^\p} - S_{21^\p}\right]\right\}.\label{Real14}\\
\frac{\dd \sigma^{L}_{2\times 3}}{\dd^2 \bk\, \dd y} = & 
- 4 \frac{e^2 Q^2 g^2 N_c}{(2\pi)^{6}} \int \dd z_1 \dd z_2 \,
\delta ( 1 - z_1 - z_2 - z) \,
\frac{z_2}{z} z_1 (1- z_1) 
\left[z_1 (1 - z_1) + z_2 (1 - z_2)\right] \nonumber\\
&
\frac{1}{(2 \pi)^2}  
\int \dd^{8} \bx  \, 
\frac{\bx_{2^\p2}\cdot\bx_{3^\p 1}}{\bx_{2^\p2}^2 \bx_{3^\p 1}^2}\,
\left[K_0(|\bx_{12}|Q_1) \, K_0(Q X^\p)\right]_{1^\p = 1} \,
e^{i \bk \cdot\left[\bx_{3^\p 2} +\frac{z_2}{z} \bx_{2^\p2}\right]}
\nonumber \\
&
\left\{\frac{N_c}{2} \left[S_{3^\p2^\p} S_{23^\p} - S_{3^\p2^\p} S_{13^\p}\right] 
+ C_F \left[1 - S_{12}\right] - \frac{1}{2 N_c} \left[S_{22^\p} - S_{2^\p1}\right]\right\}.\label{Real23}\\
\frac{\dd \sigma^{L}_{2\times 4}}{\dd^2 \bk\, \dd y} = & 
4 \frac{e^2 Q^2 g^2 N_c}{(2\pi)^{6}} \int \dd z_1 \dd z_2 \,
\delta ( 1 - z_1 - z_2 - z) \,
\frac{z_1}{z} z_1 (1- z_1) \left[(1 - z_1)^2 + z_2^2\right] \nonumber \\
&
\frac{1}{(2 \pi)^2}  
\int \dd^8 \bx \, 
\frac{\bx_{2^\p2}\cdot\bx_{3^\p 2^\p}}{\bx_{2^\p2}^2 \bx_{3^\p 2^\p}^2}\,
\left[K_0(|\bx_{12}|Q_1) \, K_0(Q X^\p)\right]_{1^\p = 1} \,
e^{i \bk \cdot\left[\bx_{3^\p 2} +\frac{z_2}{z} \bx_{2^\p2}\right]}
\nonumber \\
&
\left\{\frac{N_c}{2} \left[S_{2^\p3^\p} S_{23^\p} - S_{3^\p2^\p} S_{13^\p}\right] 
+ C_F 
\left[1 - S_{12}\right] - \frac{1}{2 N_c} \left[S_{22^\p} - S_{2^\p1}\right]\right\}
\label{eq:twoxfour}
 , 
\end{align}
where $\bk$ and $y$ are the transverse momentum and rapidity of the produced gluon 
with $\dd y = \frac{\dd z}{ z}$ and $z$ is the fraction of the photon momentum carried 
by the radiated gluon $z \equiv \frac{k^+}{l^+}$. We also use the following notation 
for coordinate dependence of some of the Bessel functions 

\begin{align}
X &= \sqrt{z_1 z_2 \, \bx_{12}^2 + z_1 z \, \bx_{13}^2 + z_2 z \, \bx_{23}^2}
\nonumber \\
\end{align}

\noindent  
Furthermore, $X^\prime$ that appears in some of the terms is the 
same as $X$ above but with $\bx_1, \bx_2, \bx_3 \to \bx_{1^\p}, \bx_{2^\p}, \bx_{3^\p}$.
Expressions in Eqs. (\ref{eq:onexone})-(\ref{eq:twoxfour}) are our main result. 
Nevertheless they contain divergences which must be regulated before they can give 
meaningful results. Finally, it is worth noticing presence of a quadrupole ($S_{122^\p1^\p}$) 
in (\ref{eq:onextwo}) which is however suppressed at large $N_c$.

\section{Divergences}
Divergences can arise due to quantum corrections to tree level cross sections. 
Possible divergences include ultraviolet (UV) divergences which happen when
loop $4$-momenta go to infinity $\bk^\mu \to \infty$, soft divergences which arise
when loop $4$-momenta go to zero $\bk^\mu \to 0$ (for massless theories), collinear
divergences which arise when the angle between a radiated parton and its parent
(after radiation) go to zero ($\theta \to 0$), and rapidity divergences when the radiated
parton carries very small fraction $z$ of its parents momentum ($z \to 0$). 

It is straightforward to check that there are no UV or soft divergences in Eqs. 
(\ref{eq:onexone})-(\ref{eq:twoxfour}). UV divergences may occur when transverse separations 
(in coordinate space) go to zero, however in this limit the radiation kernels 
are UV finite as can be seen for example in
\begin{align}
\frac{\bx_{2^\p2}\cdot\bx_{3^\p 1}}{\bx_{2^\p2}^2 \bx_{3^\p 1}^2} =
- \frac{1}{2} 
\left[
\frac{(\bx_{2^\p2} - \bx_{3^\p1})^2}{\bx_{2^\p2}^2 \bx_{3^\p1}^2} 
- \frac{1}{\bx_{2^\p2}^2} - \frac{1}{\bx_{3^\p1}^2}
\right]
\end{align}
as $\bx_{2^\p} \leftrightarrow \bx_2$, $\bx_{3^\p} \leftrightarrow \bx_1$. Possible 
soft divergences (when transverse coordinates go to infinity in coordinate space) vanish due to behavior of Bessel functions at large values of their arguments or by the rapidly
oscillating phase factors. Furthermore as the gluon is produced at 
finite momentum there cannot be rapidity divergences. However, there is a collinear divergence present in some of the terms, namely when the multiple scatterings on the target happen before radiation of the gluon (Eqs. \ref{eq:onexone}), (\ref{eq:twoxtwo}). This collinear divergence appears as a pole in
\begin{align}
\frac{1}{\left[z \bp - z_1 \bk\right]^2}
\end{align} 
when integrating over the quark transverse momentum $\bp$ (and similarly for antiquark).
In principle one absorbs this collinear divergence into bare fragmentation function which 
makes it scale dependent. This was discussed in great detail in \cite{Bergabo:2022tcu} for 
contributions of quark-quark splitting $P_{qq}$ to the evolution of quark-hadron fragmentation 
function. It is worth remembering DGLAP evolution equation for scale dependence of 
quark-hadron fragmentation function is a matrix equation which mixes quark-hadron and gluon-hadron fragmentation functions. This is symbolically written as 

\begin{align}
\frac{\partial}{\partial\, log Q^2} 
\begin{pmatrix}
D_{h/q} \\
D_{h/g} 
\end{pmatrix}
\sim 
\begin{pmatrix}
P_{qq} & P_{gq} \\
P_{qg} & P_{gg}
\end{pmatrix}
\begin{pmatrix}
D_{h/q} \\
D_{h/g} 
\end{pmatrix}
\end{align}  
where in this work we are including the contribution of $P_{gq}$ channel to evolution of the 
quark-hadron fragmentation function. To do this we define a bare gluon-hadron 
fragmentation function $D_{h/g}^{(0)} (z_h)$ and convolute the partonic production cross 
section with this fragmentation function to describe hadronization of the parton. 
To be specific we focus on
Eq. (\ref{eq:onexone}); we first shift $z_1 \longrightarrow z_1 - z$ and then $\bp \longrightarrow \bp - \bk$ to get
\begin{align}
k^+ \frac{\dd \sigma_{1\times 1}}{\dd^2 \bk\, \dd k^+} = &
4 \frac{e^2 Q^2 g^2 N_c}{(2\pi)^{6}}\,  C_F 
\int \dd^{2} \bx_1 \, \dd^{2} \bx_{1^\p} \, \dd^{2} \bx_2 \,
\left[ S_{1 1^\p} - S_{12} - S_{1^\p 2} + 1\right] 
\int_z^1 \dd z_1 \, K_0(|\bx_{12}|Q_1) K_0(|\bx_{1^\p 2}|Q_1) 
\nonumber\\
&
\frac{z_1^2}{z^2} \,  z_1^2 (1 - z_1)^2 \left[1 + \left(1 - \frac{z}{z_1}\right)^2\right]
\int \dtwo{\bp} \frac{e^{i\bp\cdot\bx_{1^\p 1}}}
{\left[\bp - \frac{z_1}{z} \bk\right]^2} 
\end{align}
To go from the partonic to hadronic production cross section we convolute the parton
level cross section with the bare gluon-hadron fragmentation function 
$ D^{(0)}_{h/g} (z_h)$
\begin{align}
k^+_h \frac{\dd \sigma_{1\times 1}}{\dd^2 \bk_h \, \dd k^+_h} \equiv
\int_0^1 \frac{\dd z_h}{z_h^2} \, D^{(0)}_{h/g} (z_h) \, 
k^+\frac{\dd \sigma_{1\times 1}}{\dd^2 \bk\, \dd k^+} 
\end{align}
where $\bk_h, k^+_h$ are now the produced hadron momenta and 
$z_h \equiv \frac{k_h^\mu}{k^\mu}$. We further define a
new variable $\xi \equiv \frac{z}{z_1}$ in terms of which the hadron production
cross section can be written as
\begin{align}
k^+_h \frac{\dd \sigma_{1\times 1}}{\dd^2 \bk_h \, \dd k^+_h} =&
4 \frac{e^2 Q^2 g^2 N_c}{(2\pi)^{6}}\,  C_F 
\int \dd^{2} \bx_1 \, \dd^{2} \bx_{1^\p} \, \dd^{2} \bx_2 \,
\left[ S_{1 1^\p} - S_{12} - S_{1^\p 2} + 1\right] 
\int_0^1 \frac{\dd z_h}{z_h^2} \, D^{(0)}_{h/g} (z_h) 
\nonumber\\
&
\int_z^1 \frac{\dd \xi}{\xi^2} \, \left(\frac{z}{\xi}\right)^3 \,
\left(1 - \frac{z}{\xi}\right)^2 \, \frac{[1 + (1 - \xi)^2]}{\xi} \, 
K_0(|\bx_{12}|Q_1) K_0(|\bx_{1^\p 2}|Q_1) 
\int \dtwo{\bp} \frac{e^{i\bp\cdot\bx_{1^\p 1}}}
{\left[\bp - \frac{1}{\xi} \bk\right]^2} 
\end{align}
note that we now have $z \equiv \frac{k^+}{l^+} = \frac{k^+_h}{z_h l^+}$
and that $Q_1^2 = Q^2 \, z_1 (1 - z_1) = Q^2 \frac{z}{\xi} (1 - \frac{z}{\xi})$.
The integration over $\bp$ is divergent and must be regulated. Here we use 
dimensional regularization which gives
\begin{align}
\int \dtwo{\bp} \frac{e^{i\bp\cdot\bx_{1^\p 1}}}
{\left[\bp - \frac{1}{\xi} \bk\right]^2} =
\frac{1}{2 \pi} \, \left[
\frac{1}{\eps} - \log (\pi e^{\gamma_E} \mu \, |\bx_{1^\p1}|)
\right]
e^{i \frac{\bk_h}{\xi z_h}\cdot\bx_{1^\p1}}
\end{align}

Further defining $z^\p_h = \xi \, z_h$ and then relabeling $z^\p_h$ as $z_h$ 
and multiplying by $2$ to take into account contribution of antiquark radiation 
given by Eq. (\ref{eq:twoxtwo}) leads to
\begin{align}
k^+_h \frac{\dd \sigma_{(1\times 1 + 2\times 2)}}{\dd^2 \bk_h \, \dd k^+_h} =&
8 \frac{e^2 Q^2 N_c}{(2\pi)^{5}}\, 
\int \dd^{2} \bx_1 \, \dd^{2} \bx_{1^\p} \, \dd^{2} \bx_2 \,
\left[ S_{1 1^\p} - S_{12} - S_{1^\p 2} + 1\right] 
\int_0^1 \frac{\dd z_h}{z_h^2} \, 
\left(\frac{k^+_h}{l^+}\frac{1}{z_h}\right)^3 \,
\left(1 - \frac{k^+_h}{l^+}\frac{1}{z_h}\right)^2 \,
\nonumber\\
&
e^{i \frac{\bk_h}{z_h}\cdot\bx_{1^\p1}} \, 
K_0 (|\bx_{12}|Q_1) K_0 (|\bx_{1^\p 2}|Q_1)\, 
\int_{z_h}^1 \frac{\dd \xi}{\xi} \,
\frac{\alpha_s}{\pi} \,  P_{gq} (\xi) \, 
\left[
\frac{1}{\eps} - \log (\pi e^{\gamma_E} \mu \, |\bx_{1^\p1}|)
\right] \, D^{(0)}_{h/g} (\frac{z_h}{\xi}) 
\end{align}
where we now have $Q_1^2 = Q^2 \frac{k^+_h}{z_h l^+}  (1 - \frac{k^+_h}{z_h l^+})$
and the quark-gluon splitting function $P_{gq}$ is defined as
\begin{align}
P_{gq} (\xi) \equiv C_F \, \frac{1 + (1 - \xi)^2}{\xi}
\end{align}
This is our result for Eqs. (\ref{eq:onexone}) and (\ref{eq:twoxtwo}) which are
the only terms containing divergences. We note that unlike the quark-quark splitting
which requires a $+$ prescription to regulate the large $\xi$ limit there is no need 
for such a prescription here in this channel. Such a prescription would have come 
from the would be virtual corrections which do not exist to this order in this channel.

At this point one can add the contribution of the LO term as well as the case of 
quark-quark splitting channel already considered 
in~\cite{Bergabo:2022zhe}, the full result for terms involving collinear divergences is 
\begin{align}
k^+_h \frac{\dd \sigma}{\dd^2 \bk_h \, \dd k^+_h} =&
8 \frac{e^2 Q^2 N_c}{(2\pi)^{5}}\, 
\int \dd^{2} \bx_1 \, \dd^{2} \bx_{1^\p} \, \dd^{2} \bx_2 \,
\left[ S_{1 1^\p} - S_{12} - S_{1^\p 2} + 1\right] 
\int_0^1 \frac{\dd z_h}{z_h^2} \, 
\left(\frac{k^+_h}{l^+}\frac{1}{z_h}\right)^3 \,
\left(1 - \frac{k^+_h}{l^+}\frac{1}{z_h}\right)^2 \,
\nonumber\\
&
e^{i \frac{\bk_h}{z_h}\cdot\bx_{1^\p1}}
K_0 (|\bx_{12}|Q_1) K_0 (|\bx_{1^\p 2}|Q_1)\, 
\int_{z_h}^1 \frac{\dd \xi}{\xi} \,
\Bigg\{
\Bigg[\delta (1 - \xi) + 
\frac{\alpha_s}{\pi} \,  P_{qq} (\xi) \, 
\left[\frac{1}{\eps} - \log (\pi e^{\gamma_E} \mu \, |\bx_{1^\p1}|)\right] 
\, D^{(0)}_{h/q} (\frac{z_h}{\xi})
\Bigg]
\nonumber \\
&
+
\frac{\alpha_s}{\pi} \,  P_{gq} (\xi) \, 
\left[
\frac{1}{\eps} - \log (\pi e^{\gamma_E} \mu \, |\bx_{1^\p1}|)
\right] \, D^{(0)}_{h/g} (\frac{z_h}{\xi}) 
\Bigg\}
\end{align}
It is standard to subtract the $1/\eps$ divergence in addition to 
some/no finite terms ($\gamma_E , \cdots$) depending on the renormalization 
scheme (MS, $\overline{MS}, \cdots$) or to cancel them using 
counter terms and to define the renormalized parton-hadron fragmentation 
function $D (z_h, \mu^2)$ in terms of which the single inclusive hadron 
production cross section can be written as
\begin{align}
k^+_h \frac{\dd \sigma^L_{SIDIS}}{\dd^2 \bk_h \, \dd k^+_h} =&
8 \frac{e^2 Q^2 N_c}{(2\pi)^{5}}\, 
\int \dd^{2} \bx_1 \, \dd^{2} \bx_{1^\p} \, \dd^{2} \bx_2 \,
\left[ S_{1 1^\p} - S_{12} - S_{1^\p 2} + 1\right] 
\int_0^1 \frac{\dd z_h}{z_h^2} \, 
\left(\frac{k^+_h}{l^+}\frac{1}{z_h}\right)^3 \,
\left(1 - \frac{k^+_h}{l^+}\frac{1}{z_h}\right)^2 
\nonumber\\
&
e^{i \frac{\bk_h}{z_h}\cdot\bx_{1^\p1}} \, 
K_0 (|\bx_{12}|Q_1) K_0 (|\bx_{1^\p 2}|Q_1)\, D_{h/q} (z_h, \mu^2) 
+ \cdots .
\end{align}
where $D_{h/q} (z_h, \mu^2)$ now includes contribution of both quark-quark and quark-gluon 
splitting channels and $\cdots$ represents the finite terms given in 
Eqs. (\ref{eq:onextwo})-(\ref{eq:twoxfour}) for gluons (and analogous but different terms 
for quarks~\cite{Bergabo:2022zhe}). 

It will be interesting to investigate the properties of these results in various limits; for 
example when the produced hadron transverse momentum $|\bk|$ is much smaller than the photon 
virtuality $Q$ one expects to encounter the so-called Sudakov logs ($\log Q^2/\bk^2$) 
which could be large and hence would require resummation. This would further complicate the
search for saturation effects as, depending on kinematics, they could be masked by Sudakov 
effects. A possible way around this would be to measure single inclusive hadrons with 
$\bk^2 \sim Q^2$ to minimize the effect of Sudakov logs.

It is worth mentioning that if one further integrates over the produced 
gluon transverse momentum $\bk$ one can obtain the rapidity dependence of parton multiplicity distribution 
$\frac{\dd N_g}{d y}$~\cite{Caucal:2024cdq}. This would then require a mechanism besides 
fragmentation functions to describe hadronization of the produced gluons. This may 
require some care specially for the collinear divergent terms in Eqs. 
(\ref{eq:onexone})-(\ref{eq:twoxtwo}). This is however beyond the score of this work and will be investigated later.

In summary we have calculated contribution of gluons to single inclusive 
hadron production in Deep Inelastic Scattering at small $x$ and at finite $N_c$
for longitudinal photon exchange. These contributions
are part of the next-to-leading order corrections to SIDIS. Some of these
contributions exhibit collinear divergences which are then absorbed into 
Leading Log evolution of fragmentations functions. The rest are finite and 
constitute part of NLO contributions to SIDIS. The case of transverse photon 
exchange is conceptually similar but algebraically more tedious and will be reported 
elsewhere.

Acknowledgements: 
This material is based upon work supported by the U.S. Department of Energy, Office of Science, Office of Nuclear Physics, within the framework of the Saturated Glue (SURGE) Topical Theory Collaboration. We gratefully acknowledge support from the DOE Office of Nuclear Physics through Grant No. DE-SC0002307. We would like to thank T. Altinoluk, 
N. Armesto, G. Beuf, E. Iancu, Y. Kovchegov, T. Lappi, C. Marquet, F. Salazar and B. Xiao for helpful discussions. J.J-M would like to thank University of Santiago de Compostela,  
Instituto Galego de Física de Altas Enerxías (IGFAE), Jyvaskyla University, Ecole Polytechnique CPHT, Universite Paris-Saclay IPHT and CERN for hospitality and financial support.

\begin{appendices}

\section{Structure of finite $N_c$ correlations}

In this work we have kept the $N_c$ suppressed parts of NLO corrections while in
\cite{Bergabo:2022tcu,Bergabo:2023wed,Bergabo:2022zhe} we made the large $N_c$ approximation. Rather than repeating the full results derived in those papers keeping $N_c$ finite we show the finite $N_c$ generalization of those expressions for the case
of {\it dihadron} production. To get the finite $N_c$ results for the case of quark hadronizing as done in \cite{Bergabo:2022zhe} one would integrate over the final state phase space for antiquark and gluon which would then set some of these coordinates equal. Labeling of terms corresponds to the diagrams involved as before (virtual diagrams are shown in \cite{Bergabo:2022tcu,Bergabo:2023wed,Bergabo:2022zhe}). We also factor out an overall factor of $N_c$ since the LO result also contains an explicit $N_c$. These expressions then can be used to replace the large $N_c$ expressions in \cite{Bergabo:2022tcu,Bergabo:2023wed,Bergabo:2022zhe} easily.

\bea
\sigma_{1\times 1} &=& \sigma_{2\times 2} = 
 N_c\, C_F \left\{ S_{122^\p1^\p} - S_{12} - S_{2^\p1^\p} + 1\right\}
\\
\sigma_{1\times 2} &= & N_c \, 
\left\{\frac{N_c}{2} S_{12}S_{2^\p1^\p} + C_F \left[1 - S_{12} - S_{2^\p1^\p}\right] - \frac{1}{2 N_c} S_{122^\p1^\p}\right\}
\\
\sigma_{3\times 3} &=& \sigma_{4\times 4} = \sigma_{3\times 4} = N_c \, 
\left\{\frac{N_c}{2} \left[S_{133^\p1^\p} S_{322^\p3^\p} - S_{13} S_{32} - S_{3^\p1^\p} S_{2^\p3^\p}\right] 
+ C_F - \frac{1}{2 N_c} \left[S_{122^\p1^\p} - S_{12} - S_{2^\p1^\p} \right] \right\}
\\
\sigma_{1\times 3} &=& \sigma_{1\times 4} = N_c \, 
\left\{\frac{N_c}{2} \left[S_{122^\p3^\p} S_{3^\p1^\p} - S_{3^\p1^\p} S_{2^\p3^\p}\right] 
+ C_F \left[1 - S_{12}\right] - \frac{1}{2 N_c} \left[S_{122^\p1^\p} - S_{2^\p1^\p}\right]\right\}
\\
\sigma_{2\times 3} &=& \sigma_{2\times 4} = N_c \, 
\left\{\frac{N_c}{2} \left[S_{123^\p1^\p} S_{2^\p3^\p} - S_{3^\p1^\p} S_{2^\p3^\p}\right] 
+ C_F \left[1 - S_{12}\right] - \frac{1}{2 N_c} \left[S_{122^\p1^\p} - S_{2^\p1^\p}\right]\right\}
\\
\sigma_{5 \times LO} &=& \sigma_{7\times LO} = N_c \, 
\left\{\frac{N_c}{2} \left[S_{322^\p1^\p} S_{13} - S_{32} S_{13}\right] 
+ C_F \left[1 - S_{2^\p1^\p}\right] - \frac{1}{2 N_c} \left[S_{122^\p1^\p} - S_{12}\right]\right\}
\\
\sigma_{6 \times LO} &=& \sigma_{8\times LO} = N_c \, 
\left\{\frac{N_c}{2} \left[S_{2^\p1^\p13} S_{32} - S_{13} S_{32}\right] 
+ C_F \left[1 - S_{2^\p1^\p}\right] - \frac{1}{2 N_c} \left[S_{122^\p1^\p} - S_{12}\right]\right\}
\\
\sigma_{9 \times LO} &=& \sigma_{10\times LO} = \sigma_{11\times LO} = \sigma_{12\times LO} = \sigma_{14\times LO} = 
N_c\, C_F \left\{ S_{122^\p1^\p} - S_{12} - S_{2^\p1^\p} + 1\right\}
\\
\sigma_{13\times LO} &= & N_c \, 
\left\{\frac{N_c}{2} \left[S_{2^\p1^\p} S_{12} \right] 
+ C_F \left[1 - S_{2^\p1^\p} - S_{12}\right] - \frac{1}{2 N_c} \left[S_{122^\p1^\p}\right]\right\}
\eea

\end{appendices}

\bibliography{mybib}

\begin{thebibliography}{133}
\expandafter\ifx\csname natexlab\endcsname\relax\def\natexlab#1{#1}\fi
\expandafter\ifx\csname bibnamefont\endcsname\relax
  \def\bibnamefont#1{#1}\fi
\expandafter\ifx\csname bibfnamefont\endcsname\relax
  \def\bibfnamefont#1{#1}\fi
\expandafter\ifx\csname citenamefont\endcsname\relax
  \def\citenamefont#1{#1}\fi
\expandafter\ifx\csname url\endcsname\relax
  \def\url#1{\texttt{#1}}\fi
\expandafter\ifx\csname urlprefix\endcsname\relax\def\urlprefix{URL }\fi
\providecommand{\bibinfo}[2]{#2}
\providecommand{\eprint}[2][]{\url{#2}}

\bibitem[{\citenamefont{Aschenauer et~al.}(2016)}]{Aschenauer:2016our}
\bibinfo{author}{\bibfnamefont{E.-C.} \bibnamefont{Aschenauer}}
  \bibnamefont{et~al.} (\bibinfo{year}{2016}), \eprint{1602.03922}.

\bibitem[{\citenamefont{Accardi et~al.}(2016)}]{Accardi:2012qut}
\bibinfo{author}{\bibfnamefont{A.}~\bibnamefont{Accardi}} \bibnamefont{et~al.},
  \bibinfo{journal}{Eur. Phys. J. A} \textbf{\bibinfo{volume}{52}},
  \bibinfo{pages}{268} (\bibinfo{year}{2016}), \eprint{1212.1701}.

\bibitem[{\citenamefont{Gribov et~al.}(1983)\citenamefont{Gribov, Levin, and
  Ryskin}}]{Gribov:1983ivg}
\bibinfo{author}{\bibfnamefont{L.~V.} \bibnamefont{Gribov}},
  \bibinfo{author}{\bibfnamefont{E.~M.} \bibnamefont{Levin}}, \bibnamefont{and}
  \bibinfo{author}{\bibfnamefont{M.~G.} \bibnamefont{Ryskin}},
  \bibinfo{journal}{Phys. Rept.} \textbf{\bibinfo{volume}{100}},
  \bibinfo{pages}{1} (\bibinfo{year}{1983}).

\bibitem[{\citenamefont{Mueller and Qiu}(1986)}]{Mueller:1985wy}
\bibinfo{author}{\bibfnamefont{A.~H.} \bibnamefont{Mueller}} \bibnamefont{and}
  \bibinfo{author}{\bibfnamefont{J.-w.} \bibnamefont{Qiu}},
  \bibinfo{journal}{Nucl. Phys. B} \textbf{\bibinfo{volume}{268}},
  \bibinfo{pages}{427} (\bibinfo{year}{1986}).

\bibitem[{\citenamefont{McLerran and
  Venugopalan}(1994{\natexlab{a}})}]{McLerran:1993ni}
\bibinfo{author}{\bibfnamefont{L.~D.} \bibnamefont{McLerran}} \bibnamefont{and}
  \bibinfo{author}{\bibfnamefont{R.}~\bibnamefont{Venugopalan}},
  \bibinfo{journal}{Phys. Rev. D} \textbf{\bibinfo{volume}{49}},
  \bibinfo{pages}{2233} (\bibinfo{year}{1994}{\natexlab{a}}),
  \eprint{hep-ph/9309289}.

\bibitem[{\citenamefont{McLerran and
  Venugopalan}(1994{\natexlab{b}})}]{McLerran:1993ka}
\bibinfo{author}{\bibfnamefont{L.~D.} \bibnamefont{McLerran}} \bibnamefont{and}
  \bibinfo{author}{\bibfnamefont{R.}~\bibnamefont{Venugopalan}},
  \bibinfo{journal}{Phys. Rev. D} \textbf{\bibinfo{volume}{49}},
  \bibinfo{pages}{3352} (\bibinfo{year}{1994}{\natexlab{b}}),
  \eprint{hep-ph/9311205}.

\bibitem[{\citenamefont{Jalilian-Marian
  et~al.}(1997{\natexlab{a}})\citenamefont{Jalilian-Marian, Kovner, McLerran,
  and Weigert}}]{Jalilian-Marian:1996mkd}
\bibinfo{author}{\bibfnamefont{J.}~\bibnamefont{Jalilian-Marian}},
  \bibinfo{author}{\bibfnamefont{A.}~\bibnamefont{Kovner}},
  \bibinfo{author}{\bibfnamefont{L.~D.} \bibnamefont{McLerran}},
  \bibnamefont{and} \bibinfo{author}{\bibfnamefont{H.}~\bibnamefont{Weigert}},
  \bibinfo{journal}{Phys. Rev. D} \textbf{\bibinfo{volume}{55}},
  \bibinfo{pages}{5414} (\bibinfo{year}{1997}{\natexlab{a}}),
  \eprint{hep-ph/9606337}.

\bibitem[{\citenamefont{Jalilian-Marian and
  Kovchegov}(2004)}]{Jalilian-Marian:2004vhw}
\bibinfo{author}{\bibfnamefont{J.}~\bibnamefont{Jalilian-Marian}}
  \bibnamefont{and} \bibinfo{author}{\bibfnamefont{Y.~V.}
  \bibnamefont{Kovchegov}}, \bibinfo{journal}{Phys. Rev. D}
  \textbf{\bibinfo{volume}{70}}, \bibinfo{pages}{114017}
  (\bibinfo{year}{2004}), \bibinfo{note}{[Erratum: Phys.Rev.D 71, 079901
  (2005)]}, \eprint{hep-ph/0405266}.

\bibitem[{\citenamefont{Dumitru et~al.}(2006)\citenamefont{Dumitru,
  Hayashigaki, and Jalilian-Marian}}]{Dumitru:2005gt}
\bibinfo{author}{\bibfnamefont{A.}~\bibnamefont{Dumitru}},
  \bibinfo{author}{\bibfnamefont{A.}~\bibnamefont{Hayashigaki}},
  \bibnamefont{and}
  \bibinfo{author}{\bibfnamefont{J.}~\bibnamefont{Jalilian-Marian}},
  \bibinfo{journal}{Nucl. Phys. A} \textbf{\bibinfo{volume}{765}},
  \bibinfo{pages}{464} (\bibinfo{year}{2006}), \eprint{hep-ph/0506308}.

\bibitem[{\citenamefont{Jalilian-Marian}(2006)}]{Jalilian-Marian:2005qbq}
\bibinfo{author}{\bibfnamefont{J.}~\bibnamefont{Jalilian-Marian}},
  \bibinfo{journal}{Nucl. Phys. A} \textbf{\bibinfo{volume}{770}},
  \bibinfo{pages}{210} (\bibinfo{year}{2006}), \eprint{hep-ph/0509338}.

\bibitem[{\citenamefont{Marquet}(2007)}]{Marquet:2007vb}
\bibinfo{author}{\bibfnamefont{C.}~\bibnamefont{Marquet}},
  \bibinfo{journal}{Nucl. Phys. A} \textbf{\bibinfo{volume}{796}},
  \bibinfo{pages}{41} (\bibinfo{year}{2007}), \eprint{0708.0231}.

\bibitem[{\citenamefont{Albacete and Marquet}(2010)}]{Albacete:2010pg}
\bibinfo{author}{\bibfnamefont{J.~L.} \bibnamefont{Albacete}} \bibnamefont{and}
  \bibinfo{author}{\bibfnamefont{C.}~\bibnamefont{Marquet}},
  \bibinfo{journal}{Phys. Rev. Lett.} \textbf{\bibinfo{volume}{105}},
  \bibinfo{pages}{162301} (\bibinfo{year}{2010}), \eprint{1005.4065}.

\bibitem[{\citenamefont{Stasto et~al.}(2012)\citenamefont{Stasto, Xiao, and
  Yuan}}]{Stasto:2011ru}
\bibinfo{author}{\bibfnamefont{A.}~\bibnamefont{Stasto}},
  \bibinfo{author}{\bibfnamefont{B.-W.} \bibnamefont{Xiao}}, \bibnamefont{and}
  \bibinfo{author}{\bibfnamefont{F.}~\bibnamefont{Yuan}},
  \bibinfo{journal}{Phys. Lett. B} \textbf{\bibinfo{volume}{716}},
  \bibinfo{pages}{430} (\bibinfo{year}{2012}), \eprint{1109.1817}.

\bibitem[{\citenamefont{Lappi and Mantysaari}(2013)}]{Lappi:2012nh}
\bibinfo{author}{\bibfnamefont{T.}~\bibnamefont{Lappi}} \bibnamefont{and}
  \bibinfo{author}{\bibfnamefont{H.}~\bibnamefont{Mantysaari}},
  \bibinfo{journal}{Nucl. Phys. A} \textbf{\bibinfo{volume}{908}},
  \bibinfo{pages}{51} (\bibinfo{year}{2013}), \eprint{1209.2853}.

\bibitem[{\citenamefont{Jalilian-Marian and
  Rezaeian}(2012{\natexlab{a}})}]{Jalilian-Marian:2012wwi}
\bibinfo{author}{\bibfnamefont{J.}~\bibnamefont{Jalilian-Marian}}
  \bibnamefont{and} \bibinfo{author}{\bibfnamefont{A.~H.}
  \bibnamefont{Rezaeian}}, \bibinfo{journal}{Phys. Rev. D}
  \textbf{\bibinfo{volume}{86}}, \bibinfo{pages}{034016}
  (\bibinfo{year}{2012}{\natexlab{a}}), \eprint{1204.1319}.

\bibitem[{\citenamefont{Jalilian-Marian and
  Rezaeian}(2012{\natexlab{b}})}]{Jalilian-Marian:2011tvq}
\bibinfo{author}{\bibfnamefont{J.}~\bibnamefont{Jalilian-Marian}}
  \bibnamefont{and} \bibinfo{author}{\bibfnamefont{A.~H.}
  \bibnamefont{Rezaeian}}, \bibinfo{journal}{Phys. Rev. D}
  \textbf{\bibinfo{volume}{85}}, \bibinfo{pages}{014017}
  (\bibinfo{year}{2012}{\natexlab{b}}), \eprint{1110.2810}.

\bibitem[{\citenamefont{Zheng et~al.}(2014)\citenamefont{Zheng, Aschenauer,
  Lee, and Xiao}}]{Zheng:2014vka}
\bibinfo{author}{\bibfnamefont{L.}~\bibnamefont{Zheng}},
  \bibinfo{author}{\bibfnamefont{E.~C.} \bibnamefont{Aschenauer}},
  \bibinfo{author}{\bibfnamefont{J.~H.} \bibnamefont{Lee}}, \bibnamefont{and}
  \bibinfo{author}{\bibfnamefont{B.-W.} \bibnamefont{Xiao}},
  \bibinfo{journal}{Phys. Rev. D} \textbf{\bibinfo{volume}{89}},
  \bibinfo{pages}{074037} (\bibinfo{year}{2014}), \eprint{1403.2413}.

\bibitem[{\citenamefont{Stasto et~al.}(2018)\citenamefont{Stasto, Wei, Xiao,
  and Yuan}}]{Stasto:2018rci}
\bibinfo{author}{\bibfnamefont{A.}~\bibnamefont{Stasto}},
  \bibinfo{author}{\bibfnamefont{S.-Y.} \bibnamefont{Wei}},
  \bibinfo{author}{\bibfnamefont{B.-W.} \bibnamefont{Xiao}}, \bibnamefont{and}
  \bibinfo{author}{\bibfnamefont{F.}~\bibnamefont{Yuan}},
  \bibinfo{journal}{Phys. Lett. B} \textbf{\bibinfo{volume}{784}},
  \bibinfo{pages}{301} (\bibinfo{year}{2018}), \eprint{1805.05712}.

\bibitem[{\citenamefont{Albacete et~al.}(2019)\citenamefont{Albacete,
  Giacalone, Marquet, and Matas}}]{Albacete:2018ruq}
\bibinfo{author}{\bibfnamefont{J.~L.} \bibnamefont{Albacete}},
  \bibinfo{author}{\bibfnamefont{G.}~\bibnamefont{Giacalone}},
  \bibinfo{author}{\bibfnamefont{C.}~\bibnamefont{Marquet}}, \bibnamefont{and}
  \bibinfo{author}{\bibfnamefont{M.}~\bibnamefont{Matas}},
  \bibinfo{journal}{Phys. Rev. D} \textbf{\bibinfo{volume}{99}},
  \bibinfo{pages}{014002} (\bibinfo{year}{2019}), \eprint{1805.05711}.

\bibitem[{\citenamefont{M\"antysaari et~al.}(2020)\citenamefont{M\"antysaari,
  Mueller, Salazar, and Schenke}}]{Mantysaari:2019hkq}
\bibinfo{author}{\bibfnamefont{H.}~\bibnamefont{M\"antysaari}},
  \bibinfo{author}{\bibfnamefont{N.}~\bibnamefont{Mueller}},
  \bibinfo{author}{\bibfnamefont{F.}~\bibnamefont{Salazar}}, \bibnamefont{and}
  \bibinfo{author}{\bibfnamefont{B.}~\bibnamefont{Schenke}},
  \bibinfo{journal}{Phys. Rev. Lett.} \textbf{\bibinfo{volume}{124}},
  \bibinfo{pages}{112301} (\bibinfo{year}{2020}), \eprint{1912.05586}.

\bibitem[{\citenamefont{Hatta et~al.}(2021{\natexlab{a}})\citenamefont{Hatta,
  Xiao, Yuan, and Zhou}}]{Hatta:2020bgy}
\bibinfo{author}{\bibfnamefont{Y.}~\bibnamefont{Hatta}},
  \bibinfo{author}{\bibfnamefont{B.-W.} \bibnamefont{Xiao}},
  \bibinfo{author}{\bibfnamefont{F.}~\bibnamefont{Yuan}}, \bibnamefont{and}
  \bibinfo{author}{\bibfnamefont{J.}~\bibnamefont{Zhou}},
  \bibinfo{journal}{Phys. Rev. Lett.} \textbf{\bibinfo{volume}{126}},
  \bibinfo{pages}{142001} (\bibinfo{year}{2021}{\natexlab{a}}),
  \eprint{2010.10774}.

\bibitem[{\citenamefont{Jia et~al.}(2020)\citenamefont{Jia, Wei, Xiao, and
  Yuan}}]{Jia:2019qbl}
\bibinfo{author}{\bibfnamefont{J.}~\bibnamefont{Jia}},
  \bibinfo{author}{\bibfnamefont{S.-Y.} \bibnamefont{Wei}},
  \bibinfo{author}{\bibfnamefont{B.-W.} \bibnamefont{Xiao}}, \bibnamefont{and}
  \bibinfo{author}{\bibfnamefont{F.}~\bibnamefont{Yuan}},
  \bibinfo{journal}{Phys. Rev. D} \textbf{\bibinfo{volume}{101}},
  \bibinfo{pages}{094008} (\bibinfo{year}{2020}), \eprint{1910.05290}.

\bibitem[{\citenamefont{Gelis and Jalilian-Marian}(2002)}]{Gelis:2002fw}
\bibinfo{author}{\bibfnamefont{F.}~\bibnamefont{Gelis}} \bibnamefont{and}
  \bibinfo{author}{\bibfnamefont{J.}~\bibnamefont{Jalilian-Marian}},
  \bibinfo{journal}{Phys. Rev. D} \textbf{\bibinfo{volume}{66}},
  \bibinfo{pages}{094014} (\bibinfo{year}{2002}), \eprint{hep-ph/0208141}.

\bibitem[{\citenamefont{Dominguez et~al.}(2011)\citenamefont{Dominguez,
  Marquet, Xiao, and Yuan}}]{Dominguez:2011wm}
\bibinfo{author}{\bibfnamefont{F.}~\bibnamefont{Dominguez}},
  \bibinfo{author}{\bibfnamefont{C.}~\bibnamefont{Marquet}},
  \bibinfo{author}{\bibfnamefont{B.-W.} \bibnamefont{Xiao}}, \bibnamefont{and}
  \bibinfo{author}{\bibfnamefont{F.}~\bibnamefont{Yuan}},
  \bibinfo{journal}{Phys. Rev. D} \textbf{\bibinfo{volume}{83}},
  \bibinfo{pages}{105005} (\bibinfo{year}{2011}), \eprint{1101.0715}.

\bibitem[{\citenamefont{Metz and Zhou}(2011)}]{Metz:2011wb}
\bibinfo{author}{\bibfnamefont{A.}~\bibnamefont{Metz}} \bibnamefont{and}
  \bibinfo{author}{\bibfnamefont{J.}~\bibnamefont{Zhou}},
  \bibinfo{journal}{Phys. Rev. D} \textbf{\bibinfo{volume}{84}},
  \bibinfo{pages}{051503} (\bibinfo{year}{2011}), \eprint{1105.1991}.

\bibitem[{\citenamefont{Dominguez et~al.}(2012)\citenamefont{Dominguez, Qiu,
  Xiao, and Yuan}}]{Dominguez:2011br}
\bibinfo{author}{\bibfnamefont{F.}~\bibnamefont{Dominguez}},
  \bibinfo{author}{\bibfnamefont{J.-W.} \bibnamefont{Qiu}},
  \bibinfo{author}{\bibfnamefont{B.-W.} \bibnamefont{Xiao}}, \bibnamefont{and}
  \bibinfo{author}{\bibfnamefont{F.}~\bibnamefont{Yuan}},
  \bibinfo{journal}{Phys. Rev. D} \textbf{\bibinfo{volume}{85}},
  \bibinfo{pages}{045003} (\bibinfo{year}{2012}), \eprint{1109.6293}.

\bibitem[{\citenamefont{Iancu and Laidet}(2013)}]{Iancu:2013dta}
\bibinfo{author}{\bibfnamefont{E.}~\bibnamefont{Iancu}} \bibnamefont{and}
  \bibinfo{author}{\bibfnamefont{J.}~\bibnamefont{Laidet}},
  \bibinfo{journal}{Nucl. Phys. A} \textbf{\bibinfo{volume}{916}},
  \bibinfo{pages}{48} (\bibinfo{year}{2013}), \eprint{1305.5926}.

\bibitem[{\citenamefont{Altinoluk
  et~al.}(2016{\natexlab{a}})\citenamefont{Altinoluk, Armesto, Beuf, and
  Rezaeian}}]{Altinoluk:2015dpi}
\bibinfo{author}{\bibfnamefont{T.}~\bibnamefont{Altinoluk}},
  \bibinfo{author}{\bibfnamefont{N.}~\bibnamefont{Armesto}},
  \bibinfo{author}{\bibfnamefont{G.}~\bibnamefont{Beuf}}, \bibnamefont{and}
  \bibinfo{author}{\bibfnamefont{A.~H.} \bibnamefont{Rezaeian}},
  \bibinfo{journal}{Phys. Lett. B} \textbf{\bibinfo{volume}{758}},
  \bibinfo{pages}{373} (\bibinfo{year}{2016}{\natexlab{a}}),
  \eprint{1511.07452}.

\bibitem[{\citenamefont{Hatta et~al.}(2016)\citenamefont{Hatta, Xiao, and
  Yuan}}]{Hatta:2016dxp}
\bibinfo{author}{\bibfnamefont{Y.}~\bibnamefont{Hatta}},
  \bibinfo{author}{\bibfnamefont{B.-W.} \bibnamefont{Xiao}}, \bibnamefont{and}
  \bibinfo{author}{\bibfnamefont{F.}~\bibnamefont{Yuan}},
  \bibinfo{journal}{Phys. Rev. Lett.} \textbf{\bibinfo{volume}{116}},
  \bibinfo{pages}{202301} (\bibinfo{year}{2016}), \eprint{1601.01585}.

\bibitem[{\citenamefont{Dumitru et~al.}(2015)\citenamefont{Dumitru, Lappi, and
  Skokov}}]{Dumitru:2015gaa}
\bibinfo{author}{\bibfnamefont{A.}~\bibnamefont{Dumitru}},
  \bibinfo{author}{\bibfnamefont{T.}~\bibnamefont{Lappi}}, \bibnamefont{and}
  \bibinfo{author}{\bibfnamefont{V.}~\bibnamefont{Skokov}},
  \bibinfo{journal}{Phys. Rev. Lett.} \textbf{\bibinfo{volume}{115}},
  \bibinfo{pages}{252301} (\bibinfo{year}{2015}), \eprint{1508.04438}.

\bibitem[{\citenamefont{Kotko et~al.}(2015)\citenamefont{Kotko, Kutak, Marquet,
  Petreska, Sapeta, and van Hameren}}]{Kotko:2015ura}
\bibinfo{author}{\bibfnamefont{P.}~\bibnamefont{Kotko}},
  \bibinfo{author}{\bibfnamefont{K.}~\bibnamefont{Kutak}},
  \bibinfo{author}{\bibfnamefont{C.}~\bibnamefont{Marquet}},
  \bibinfo{author}{\bibfnamefont{E.}~\bibnamefont{Petreska}},
  \bibinfo{author}{\bibfnamefont{S.}~\bibnamefont{Sapeta}}, \bibnamefont{and}
  \bibinfo{author}{\bibfnamefont{A.}~\bibnamefont{van Hameren}},
  \bibinfo{journal}{JHEP} \textbf{\bibinfo{volume}{09}}, \bibinfo{pages}{106}
  (\bibinfo{year}{2015}), \eprint{1503.03421}.

\bibitem[{\citenamefont{Marquet et~al.}(2016)\citenamefont{Marquet, Petreska,
  and Roiesnel}}]{Marquet:2016cgx}
\bibinfo{author}{\bibfnamefont{C.}~\bibnamefont{Marquet}},
  \bibinfo{author}{\bibfnamefont{E.}~\bibnamefont{Petreska}}, \bibnamefont{and}
  \bibinfo{author}{\bibfnamefont{C.}~\bibnamefont{Roiesnel}},
  \bibinfo{journal}{JHEP} \textbf{\bibinfo{volume}{10}}, \bibinfo{pages}{065}
  (\bibinfo{year}{2016}), \eprint{1608.02577}.

\bibitem[{\citenamefont{van Hameren et~al.}(2016)\citenamefont{van Hameren,
  Kotko, Kutak, Marquet, Petreska, and Sapeta}}]{vanHameren:2016ftb}
\bibinfo{author}{\bibfnamefont{A.}~\bibnamefont{van Hameren}},
  \bibinfo{author}{\bibfnamefont{P.}~\bibnamefont{Kotko}},
  \bibinfo{author}{\bibfnamefont{K.}~\bibnamefont{Kutak}},
  \bibinfo{author}{\bibfnamefont{C.}~\bibnamefont{Marquet}},
  \bibinfo{author}{\bibfnamefont{E.}~\bibnamefont{Petreska}}, \bibnamefont{and}
  \bibinfo{author}{\bibfnamefont{S.}~\bibnamefont{Sapeta}},
  \bibinfo{journal}{JHEP} \textbf{\bibinfo{volume}{12}}, \bibinfo{pages}{034}
  (\bibinfo{year}{2016}), \bibinfo{note}{[Erratum: JHEP 02, 158 (2019)]},
  \eprint{1607.03121}.

\bibitem[{\citenamefont{Marquet et~al.}(2018)\citenamefont{Marquet, Roiesnel,
  and Taels}}]{Marquet:2017xwy}
\bibinfo{author}{\bibfnamefont{C.}~\bibnamefont{Marquet}},
  \bibinfo{author}{\bibfnamefont{C.}~\bibnamefont{Roiesnel}}, \bibnamefont{and}
  \bibinfo{author}{\bibfnamefont{P.}~\bibnamefont{Taels}},
  \bibinfo{journal}{Phys. Rev. D} \textbf{\bibinfo{volume}{97}},
  \bibinfo{pages}{014004} (\bibinfo{year}{2018}), \eprint{1710.05698}.

\bibitem[{\citenamefont{Dumitru et~al.}(2019)\citenamefont{Dumitru, Skokov, and
  Ullrich}}]{Dumitru:2018kuw}
\bibinfo{author}{\bibfnamefont{A.}~\bibnamefont{Dumitru}},
  \bibinfo{author}{\bibfnamefont{V.}~\bibnamefont{Skokov}}, \bibnamefont{and}
  \bibinfo{author}{\bibfnamefont{T.}~\bibnamefont{Ullrich}},
  \bibinfo{journal}{Phys. Rev. C} \textbf{\bibinfo{volume}{99}},
  \bibinfo{pages}{015204} (\bibinfo{year}{2019}), \eprint{1809.02615}.

\bibitem[{\citenamefont{Dumitru and
  Jalilian-Marian}(2002{\natexlab{a}})}]{Dumitru:2001jn}
\bibinfo{author}{\bibfnamefont{A.}~\bibnamefont{Dumitru}} \bibnamefont{and}
  \bibinfo{author}{\bibfnamefont{J.}~\bibnamefont{Jalilian-Marian}},
  \bibinfo{journal}{Phys. Lett. B} \textbf{\bibinfo{volume}{547}},
  \bibinfo{pages}{15} (\bibinfo{year}{2002}{\natexlab{a}}),
  \eprint{hep-ph/0111357}.

\bibitem[{\citenamefont{Dumitru and
  Jalilian-Marian}(2002{\natexlab{b}})}]{Dumitru:2002qt}
\bibinfo{author}{\bibfnamefont{A.}~\bibnamefont{Dumitru}} \bibnamefont{and}
  \bibinfo{author}{\bibfnamefont{J.}~\bibnamefont{Jalilian-Marian}},
  \bibinfo{journal}{Phys. Rev. Lett.} \textbf{\bibinfo{volume}{89}},
  \bibinfo{pages}{022301} (\bibinfo{year}{2002}{\natexlab{b}}),
  \eprint{hep-ph/0204028}.

\bibitem[{\citenamefont{M\"antysaari et~al.}(2019)\citenamefont{M\"antysaari,
  Mueller, and Schenke}}]{Mantysaari:2019csc}
\bibinfo{author}{\bibfnamefont{H.}~\bibnamefont{M\"antysaari}},
  \bibinfo{author}{\bibfnamefont{N.}~\bibnamefont{Mueller}}, \bibnamefont{and}
  \bibinfo{author}{\bibfnamefont{B.}~\bibnamefont{Schenke}},
  \bibinfo{journal}{Phys. Rev. D} \textbf{\bibinfo{volume}{99}},
  \bibinfo{pages}{074004} (\bibinfo{year}{2019}), \eprint{1902.05087}.

\bibitem[{\citenamefont{Salazar and Schenke}(2019)}]{Salazar:2019ncp}
\bibinfo{author}{\bibfnamefont{F.}~\bibnamefont{Salazar}} \bibnamefont{and}
  \bibinfo{author}{\bibfnamefont{B.}~\bibnamefont{Schenke}},
  \bibinfo{journal}{Phys. Rev. D} \textbf{\bibinfo{volume}{100}},
  \bibinfo{pages}{034007} (\bibinfo{year}{2019}), \eprint{1905.03763}.

\bibitem[{\citenamefont{Boussarie et~al.}(2021)\citenamefont{Boussarie,
  M\"antysaari, Salazar, and Schenke}}]{Boussarie:2021ybe}
\bibinfo{author}{\bibfnamefont{R.}~\bibnamefont{Boussarie}},
  \bibinfo{author}{\bibfnamefont{H.}~\bibnamefont{M\"antysaari}},
  \bibinfo{author}{\bibfnamefont{F.}~\bibnamefont{Salazar}}, \bibnamefont{and}
  \bibinfo{author}{\bibfnamefont{B.}~\bibnamefont{Schenke}},
  \bibinfo{journal}{JHEP} \textbf{\bibinfo{volume}{09}}, \bibinfo{pages}{178}
  (\bibinfo{year}{2021}), \eprint{2106.11301}.

\bibitem[{\citenamefont{Ayala et~al.}(1996)\citenamefont{Ayala,
  Jalilian-Marian, McLerran, and Venugopalan}}]{Ayala:1995hx}
\bibinfo{author}{\bibfnamefont{A.}~\bibnamefont{Ayala}},
  \bibinfo{author}{\bibfnamefont{J.}~\bibnamefont{Jalilian-Marian}},
  \bibinfo{author}{\bibfnamefont{L.~D.} \bibnamefont{McLerran}},
  \bibnamefont{and}
  \bibinfo{author}{\bibfnamefont{R.}~\bibnamefont{Venugopalan}},
  \bibinfo{journal}{Phys. Rev. D} \textbf{\bibinfo{volume}{53}},
  \bibinfo{pages}{458} (\bibinfo{year}{1996}), \eprint{hep-ph/9508302}.

\bibitem[{\citenamefont{Jalilian-Marian}(2004)}]{Jalilian-Marian:2004cdc}
\bibinfo{author}{\bibfnamefont{J.}~\bibnamefont{Jalilian-Marian}},
  \bibinfo{journal}{Nucl. Phys. A} \textbf{\bibinfo{volume}{739}},
  \bibinfo{pages}{319} (\bibinfo{year}{2004}), \eprint{nucl-th/0402014}.

\bibitem[{\citenamefont{Kotko et~al.}(2017)\citenamefont{Kotko, Kutak, Sapeta,
  Stasto, and Strikman}}]{Kotko:2017oxg}
\bibinfo{author}{\bibfnamefont{P.}~\bibnamefont{Kotko}},
  \bibinfo{author}{\bibfnamefont{K.}~\bibnamefont{Kutak}},
  \bibinfo{author}{\bibfnamefont{S.}~\bibnamefont{Sapeta}},
  \bibinfo{author}{\bibfnamefont{A.~M.} \bibnamefont{Stasto}},
  \bibnamefont{and} \bibinfo{author}{\bibfnamefont{M.}~\bibnamefont{Strikman}},
  \bibinfo{journal}{Eur. Phys. J. C} \textbf{\bibinfo{volume}{77}},
  \bibinfo{pages}{353} (\bibinfo{year}{2017}), \eprint{1702.03063}.

\bibitem[{\citenamefont{Hagiwara et~al.}(2017)\citenamefont{Hagiwara, Hatta,
  Pasechnik, Tasevsky, and Teryaev}}]{Hagiwara:2017fye}
\bibinfo{author}{\bibfnamefont{Y.}~\bibnamefont{Hagiwara}},
  \bibinfo{author}{\bibfnamefont{Y.}~\bibnamefont{Hatta}},
  \bibinfo{author}{\bibfnamefont{R.}~\bibnamefont{Pasechnik}},
  \bibinfo{author}{\bibfnamefont{M.}~\bibnamefont{Tasevsky}}, \bibnamefont{and}
  \bibinfo{author}{\bibfnamefont{O.}~\bibnamefont{Teryaev}},
  \bibinfo{journal}{Phys. Rev. D} \textbf{\bibinfo{volume}{96}},
  \bibinfo{pages}{034009} (\bibinfo{year}{2017}), \eprint{1706.01765}.

\bibitem[{\citenamefont{Henley and Jalilian-Marian}(2006)}]{Henley:2005ms}
\bibinfo{author}{\bibfnamefont{E.~M.} \bibnamefont{Henley}} \bibnamefont{and}
  \bibinfo{author}{\bibfnamefont{J.}~\bibnamefont{Jalilian-Marian}},
  \bibinfo{journal}{Phys. Rev. D} \textbf{\bibinfo{volume}{73}},
  \bibinfo{pages}{094004} (\bibinfo{year}{2006}), \eprint{hep-ph/0512220}.

\bibitem[{\citenamefont{Klein and M\"antysaari}(2019)}]{Klein:2019qfb}
\bibinfo{author}{\bibfnamefont{S.~R.} \bibnamefont{Klein}} \bibnamefont{and}
  \bibinfo{author}{\bibfnamefont{H.}~\bibnamefont{M\"antysaari}},
  \bibinfo{journal}{Nature Rev. Phys.} \textbf{\bibinfo{volume}{1}},
  \bibinfo{pages}{662} (\bibinfo{year}{2019}), \eprint{1910.10858}.

\bibitem[{\citenamefont{Hatta et~al.}(2021{\natexlab{b}})\citenamefont{Hatta,
  Xiao, Yuan, and Zhou}}]{Hatta:2021jcd}
\bibinfo{author}{\bibfnamefont{Y.}~\bibnamefont{Hatta}},
  \bibinfo{author}{\bibfnamefont{B.-W.} \bibnamefont{Xiao}},
  \bibinfo{author}{\bibfnamefont{F.}~\bibnamefont{Yuan}}, \bibnamefont{and}
  \bibinfo{author}{\bibfnamefont{J.}~\bibnamefont{Zhou}},
  \bibinfo{journal}{Phys. Rev. D} \textbf{\bibinfo{volume}{104}},
  \bibinfo{pages}{054037} (\bibinfo{year}{2021}{\natexlab{b}}),
  \eprint{2106.05307}.

\bibitem[{\citenamefont{Kolb\'e et~al.}(2021)\citenamefont{Kolb\'e, Roy,
  Salazar, Schenke, and Venugopalan}}]{Kolbe:2020tlq}
\bibinfo{author}{\bibfnamefont{I.}~\bibnamefont{Kolb\'e}},
  \bibinfo{author}{\bibfnamefont{K.}~\bibnamefont{Roy}},
  \bibinfo{author}{\bibfnamefont{F.}~\bibnamefont{Salazar}},
  \bibinfo{author}{\bibfnamefont{B.}~\bibnamefont{Schenke}}, \bibnamefont{and}
  \bibinfo{author}{\bibfnamefont{R.}~\bibnamefont{Venugopalan}},
  \bibinfo{journal}{JHEP} \textbf{\bibinfo{volume}{01}}, \bibinfo{pages}{052}
  (\bibinfo{year}{2021}), \eprint{2008.04372}.

\bibitem[{\citenamefont{Jalilian-Marian}(2005)}]{Jalilian-Marian:2005tod}
\bibinfo{author}{\bibfnamefont{J.}~\bibnamefont{Jalilian-Marian}},
  \bibinfo{journal}{Nucl. Phys. A} \textbf{\bibinfo{volume}{753}},
  \bibinfo{pages}{307} (\bibinfo{year}{2005}), \eprint{hep-ph/0501222}.

\bibitem[{\citenamefont{Altinoluk et~al.}(2019)\citenamefont{Altinoluk,
  Boussarie, and Kotko}}]{Altinoluk:2019fui}
\bibinfo{author}{\bibfnamefont{T.}~\bibnamefont{Altinoluk}},
  \bibinfo{author}{\bibfnamefont{R.}~\bibnamefont{Boussarie}},
  \bibnamefont{and} \bibinfo{author}{\bibfnamefont{P.}~\bibnamefont{Kotko}},
  \bibinfo{journal}{JHEP} \textbf{\bibinfo{volume}{05}}, \bibinfo{pages}{156}
  (\bibinfo{year}{2019}), \eprint{1901.01175}.

\bibitem[{\citenamefont{Boussarie et~al.}(2019)\citenamefont{Boussarie,
  Grabovsky, Szymanowski, and Wallon}}]{Boussarie:2019ero}
\bibinfo{author}{\bibfnamefont{R.}~\bibnamefont{Boussarie}},
  \bibinfo{author}{\bibfnamefont{A.~V.} \bibnamefont{Grabovsky}},
  \bibinfo{author}{\bibfnamefont{L.}~\bibnamefont{Szymanowski}},
  \bibnamefont{and} \bibinfo{author}{\bibfnamefont{S.}~\bibnamefont{Wallon}},
  \bibinfo{journal}{Phys. Rev. D} \textbf{\bibinfo{volume}{100}},
  \bibinfo{pages}{074020} (\bibinfo{year}{2019}), \eprint{1905.07371}.

\bibitem[{\citenamefont{Boussarie et~al.}(2016)\citenamefont{Boussarie,
  Grabovsky, Szymanowski, and Wallon}}]{Boussarie:2016ogo}
\bibinfo{author}{\bibfnamefont{R.}~\bibnamefont{Boussarie}},
  \bibinfo{author}{\bibfnamefont{A.~V.} \bibnamefont{Grabovsky}},
  \bibinfo{author}{\bibfnamefont{L.}~\bibnamefont{Szymanowski}},
  \bibnamefont{and} \bibinfo{author}{\bibfnamefont{S.}~\bibnamefont{Wallon}},
  \bibinfo{journal}{JHEP} \textbf{\bibinfo{volume}{11}}, \bibinfo{pages}{149}
  (\bibinfo{year}{2016}), \eprint{1606.00419}.

\bibitem[{\citenamefont{Boussarie et~al.}(2014)\citenamefont{Boussarie,
  Grabovsky, Szymanowski, and Wallon}}]{Boussarie:2014lxa}
\bibinfo{author}{\bibfnamefont{R.}~\bibnamefont{Boussarie}},
  \bibinfo{author}{\bibfnamefont{A.~V.} \bibnamefont{Grabovsky}},
  \bibinfo{author}{\bibfnamefont{L.}~\bibnamefont{Szymanowski}},
  \bibnamefont{and} \bibinfo{author}{\bibfnamefont{S.}~\bibnamefont{Wallon}},
  \bibinfo{journal}{JHEP} \textbf{\bibinfo{volume}{09}}, \bibinfo{pages}{026}
  (\bibinfo{year}{2014}), \eprint{1405.7676}.

\bibitem[{\citenamefont{Dumitru and Jalilian-Marian}(2010)}]{Dumitru:2010ak}
\bibinfo{author}{\bibfnamefont{A.}~\bibnamefont{Dumitru}} \bibnamefont{and}
  \bibinfo{author}{\bibfnamefont{J.}~\bibnamefont{Jalilian-Marian}},
  \bibinfo{journal}{Phys. Rev. D} \textbf{\bibinfo{volume}{82}},
  \bibinfo{pages}{074023} (\bibinfo{year}{2010}), \eprint{1008.0480}.

\bibitem[{\citenamefont{Iancu and Venugopalan}(2003)}]{Iancu:2003xm}
\bibinfo{author}{\bibfnamefont{E.}~\bibnamefont{Iancu}} \bibnamefont{and}
  \bibinfo{author}{\bibfnamefont{R.}~\bibnamefont{Venugopalan}},
  \emph{\bibinfo{title}{{The Color glass condensate and high-energy scattering
  in QCD}}} (\bibinfo{year}{2003}), pp. \bibinfo{pages}{249--3363},
  \eprint{hep-ph/0303204}.

\bibitem[{\citenamefont{Jalilian-Marian and
  Kovchegov}(2006)}]{Jalilian-Marian:2005ccm}
\bibinfo{author}{\bibfnamefont{J.}~\bibnamefont{Jalilian-Marian}}
  \bibnamefont{and} \bibinfo{author}{\bibfnamefont{Y.~V.}
  \bibnamefont{Kovchegov}}, \bibinfo{journal}{Prog. Part. Nucl. Phys.}
  \textbf{\bibinfo{volume}{56}}, \bibinfo{pages}{104} (\bibinfo{year}{2006}),
  \eprint{hep-ph/0505052}.

\bibitem[{\citenamefont{Weigert}(2005)}]{Weigert:2005us}
\bibinfo{author}{\bibfnamefont{H.}~\bibnamefont{Weigert}},
  \bibinfo{journal}{Prog. Part. Nucl. Phys.} \textbf{\bibinfo{volume}{55}},
  \bibinfo{pages}{461} (\bibinfo{year}{2005}), \eprint{hep-ph/0501087}.

\bibitem[{\citenamefont{Gelis et~al.}(2010)\citenamefont{Gelis, Iancu,
  Jalilian-Marian, and Venugopalan}}]{Gelis:2010nm}
\bibinfo{author}{\bibfnamefont{F.}~\bibnamefont{Gelis}},
  \bibinfo{author}{\bibfnamefont{E.}~\bibnamefont{Iancu}},
  \bibinfo{author}{\bibfnamefont{J.}~\bibnamefont{Jalilian-Marian}},
  \bibnamefont{and}
  \bibinfo{author}{\bibfnamefont{R.}~\bibnamefont{Venugopalan}},
  \bibinfo{journal}{Ann. Rev. Nucl. Part. Sci.} \textbf{\bibinfo{volume}{60}},
  \bibinfo{pages}{463} (\bibinfo{year}{2010}), \eprint{1002.0333}.

\bibitem[{\citenamefont{Morreale and Salazar}(2021)}]{Morreale:2021pnn}
\bibinfo{author}{\bibfnamefont{A.}~\bibnamefont{Morreale}} \bibnamefont{and}
  \bibinfo{author}{\bibfnamefont{F.}~\bibnamefont{Salazar}},
  \bibinfo{journal}{Universe} \textbf{\bibinfo{volume}{7}},
  \bibinfo{pages}{312} (\bibinfo{year}{2021}), \eprint{2108.08254}.

\bibitem[{\citenamefont{Balitsky}(1996)}]{Balitsky:1995ub}
\bibinfo{author}{\bibfnamefont{I.}~\bibnamefont{Balitsky}},
  \bibinfo{journal}{Nucl. Phys. B} \textbf{\bibinfo{volume}{463}},
  \bibinfo{pages}{99} (\bibinfo{year}{1996}), \eprint{hep-ph/9509348}.

\bibitem[{\citenamefont{Kovchegov}(2000)}]{Kovchegov:1999yj}
\bibinfo{author}{\bibfnamefont{Y.~V.} \bibnamefont{Kovchegov}},
  \bibinfo{journal}{Phys. Rev. D} \textbf{\bibinfo{volume}{61}},
  \bibinfo{pages}{074018} (\bibinfo{year}{2000}), \eprint{hep-ph/9905214}.

\bibitem[{\citenamefont{Jalilian-Marian
  et~al.}(1997{\natexlab{b}})\citenamefont{Jalilian-Marian, Kovner, Leonidov,
  and Weigert}}]{Jalilian-Marian:1997qno}
\bibinfo{author}{\bibfnamefont{J.}~\bibnamefont{Jalilian-Marian}},
  \bibinfo{author}{\bibfnamefont{A.}~\bibnamefont{Kovner}},
  \bibinfo{author}{\bibfnamefont{A.}~\bibnamefont{Leonidov}}, \bibnamefont{and}
  \bibinfo{author}{\bibfnamefont{H.}~\bibnamefont{Weigert}},
  \bibinfo{journal}{Nucl. Phys. B} \textbf{\bibinfo{volume}{504}},
  \bibinfo{pages}{415} (\bibinfo{year}{1997}{\natexlab{b}}),
  \eprint{hep-ph/9701284}.

\bibitem[{\citenamefont{Jalilian-Marian
  et~al.}(1998{\natexlab{a}})\citenamefont{Jalilian-Marian, Kovner, Leonidov,
  and Weigert}}]{Jalilian-Marian:1997jhx}
\bibinfo{author}{\bibfnamefont{J.}~\bibnamefont{Jalilian-Marian}},
  \bibinfo{author}{\bibfnamefont{A.}~\bibnamefont{Kovner}},
  \bibinfo{author}{\bibfnamefont{A.}~\bibnamefont{Leonidov}}, \bibnamefont{and}
  \bibinfo{author}{\bibfnamefont{H.}~\bibnamefont{Weigert}},
  \bibinfo{journal}{Phys. Rev. D} \textbf{\bibinfo{volume}{59}},
  \bibinfo{pages}{014014} (\bibinfo{year}{1998}{\natexlab{a}}),
  \eprint{hep-ph/9706377}.

\bibitem[{\citenamefont{Jalilian-Marian
  et~al.}(1998{\natexlab{b}})\citenamefont{Jalilian-Marian, Kovner, and
  Weigert}}]{Jalilian-Marian:1997ubg}
\bibinfo{author}{\bibfnamefont{J.}~\bibnamefont{Jalilian-Marian}},
  \bibinfo{author}{\bibfnamefont{A.}~\bibnamefont{Kovner}}, \bibnamefont{and}
  \bibinfo{author}{\bibfnamefont{H.}~\bibnamefont{Weigert}},
  \bibinfo{journal}{Phys. Rev. D} \textbf{\bibinfo{volume}{59}},
  \bibinfo{pages}{014015} (\bibinfo{year}{1998}{\natexlab{b}}),
  \eprint{hep-ph/9709432}.

\bibitem[{\citenamefont{Jalilian-Marian
  et~al.}(1999)\citenamefont{Jalilian-Marian, Kovner, Leonidov, and
  Weigert}}]{Jalilian-Marian:1998tzv}
\bibinfo{author}{\bibfnamefont{J.}~\bibnamefont{Jalilian-Marian}},
  \bibinfo{author}{\bibfnamefont{A.}~\bibnamefont{Kovner}},
  \bibinfo{author}{\bibfnamefont{A.}~\bibnamefont{Leonidov}}, \bibnamefont{and}
  \bibinfo{author}{\bibfnamefont{H.}~\bibnamefont{Weigert}},
  \bibinfo{journal}{Phys. Rev. D} \textbf{\bibinfo{volume}{59}},
  \bibinfo{pages}{034007} (\bibinfo{year}{1999}), \bibinfo{note}{[Erratum:
  Phys.Rev.D 59, 099903 (1999)]}, \eprint{hep-ph/9807462}.

\bibitem[{\citenamefont{Kovner and Milhano}(2000)}]{Kovner:1999bj}
\bibinfo{author}{\bibfnamefont{A.}~\bibnamefont{Kovner}} \bibnamefont{and}
  \bibinfo{author}{\bibfnamefont{J.~G.} \bibnamefont{Milhano}},
  \bibinfo{journal}{Phys. Rev. D} \textbf{\bibinfo{volume}{61}},
  \bibinfo{pages}{014012} (\bibinfo{year}{2000}), \eprint{hep-ph/9904420}.

\bibitem[{\citenamefont{Kovner et~al.}(2000)\citenamefont{Kovner, Milhano, and
  Weigert}}]{Kovner:2000pt}
\bibinfo{author}{\bibfnamefont{A.}~\bibnamefont{Kovner}},
  \bibinfo{author}{\bibfnamefont{J.~G.} \bibnamefont{Milhano}},
  \bibnamefont{and} \bibinfo{author}{\bibfnamefont{H.}~\bibnamefont{Weigert}},
  \bibinfo{journal}{Phys. Rev. D} \textbf{\bibinfo{volume}{62}},
  \bibinfo{pages}{114005} (\bibinfo{year}{2000}), \eprint{hep-ph/0004014}.

\bibitem[{\citenamefont{Iancu et~al.}(2001)\citenamefont{Iancu, Leonidov, and
  McLerran}}]{Iancu:2000hn}
\bibinfo{author}{\bibfnamefont{E.}~\bibnamefont{Iancu}},
  \bibinfo{author}{\bibfnamefont{A.}~\bibnamefont{Leonidov}}, \bibnamefont{and}
  \bibinfo{author}{\bibfnamefont{L.~D.} \bibnamefont{McLerran}},
  \bibinfo{journal}{Nucl. Phys. A} \textbf{\bibinfo{volume}{692}},
  \bibinfo{pages}{583} (\bibinfo{year}{2001}), \eprint{hep-ph/0011241}.

\bibitem[{\citenamefont{Ferreiro et~al.}(2002)\citenamefont{Ferreiro, Iancu,
  Leonidov, and McLerran}}]{Ferreiro:2001qy}
\bibinfo{author}{\bibfnamefont{E.}~\bibnamefont{Ferreiro}},
  \bibinfo{author}{\bibfnamefont{E.}~\bibnamefont{Iancu}},
  \bibinfo{author}{\bibfnamefont{A.}~\bibnamefont{Leonidov}}, \bibnamefont{and}
  \bibinfo{author}{\bibfnamefont{L.}~\bibnamefont{McLerran}},
  \bibinfo{journal}{Nucl. Phys. A} \textbf{\bibinfo{volume}{703}},
  \bibinfo{pages}{489} (\bibinfo{year}{2002}), \eprint{hep-ph/0109115}.

\bibitem[{\citenamefont{Fadin and Lipatov}(1998)}]{Fadin:1998py}
\bibinfo{author}{\bibfnamefont{V.~S.} \bibnamefont{Fadin}} \bibnamefont{and}
  \bibinfo{author}{\bibfnamefont{L.~N.} \bibnamefont{Lipatov}},
  \bibinfo{journal}{Phys. Lett. B} \textbf{\bibinfo{volume}{429}},
  \bibinfo{pages}{127} (\bibinfo{year}{1998}), \eprint{hep-ph/9802290}.

\bibitem[{\citenamefont{Chirilli
  et~al.}(2012{\natexlab{a}})\citenamefont{Chirilli, Xiao, and
  Yuan}}]{Chirilli:2011km}
\bibinfo{author}{\bibfnamefont{G.~A.} \bibnamefont{Chirilli}},
  \bibinfo{author}{\bibfnamefont{B.-W.} \bibnamefont{Xiao}}, \bibnamefont{and}
  \bibinfo{author}{\bibfnamefont{F.}~\bibnamefont{Yuan}},
  \bibinfo{journal}{Phys. Rev. Lett.} \textbf{\bibinfo{volume}{108}},
  \bibinfo{pages}{122301} (\bibinfo{year}{2012}{\natexlab{a}}),
  \eprint{1112.1061}.

\bibitem[{\citenamefont{Chirilli
  et~al.}(2012{\natexlab{b}})\citenamefont{Chirilli, Xiao, and
  Yuan}}]{Chirilli:2012jd}
\bibinfo{author}{\bibfnamefont{G.~A.} \bibnamefont{Chirilli}},
  \bibinfo{author}{\bibfnamefont{B.-W.} \bibnamefont{Xiao}}, \bibnamefont{and}
  \bibinfo{author}{\bibfnamefont{F.}~\bibnamefont{Yuan}},
  \bibinfo{journal}{Phys. Rev. D} \textbf{\bibinfo{volume}{86}},
  \bibinfo{pages}{054005} (\bibinfo{year}{2012}{\natexlab{b}}),
  \eprint{1203.6139}.

\bibitem[{\citenamefont{Balitsky and
  Chirilli}(2013{\natexlab{a}})}]{balitsky:2012bs}
\bibinfo{author}{\bibfnamefont{I.}~\bibnamefont{Balitsky}} \bibnamefont{and}
  \bibinfo{author}{\bibfnamefont{G.~A.} \bibnamefont{Chirilli}},
  \bibinfo{journal}{Phys. Rev. D} \textbf{\bibinfo{volume}{87}},
  \bibinfo{pages}{014013} (\bibinfo{year}{2013}{\natexlab{a}}),
  \eprint{1207.3844}.

\bibitem[{\citenamefont{Balitsky and
  Chirilli}(2013{\natexlab{b}})}]{Balitsky:2013fea}
\bibinfo{author}{\bibfnamefont{I.}~\bibnamefont{Balitsky}} \bibnamefont{and}
  \bibinfo{author}{\bibfnamefont{G.~A.} \bibnamefont{Chirilli}},
  \bibinfo{journal}{Phys. Rev. D} \textbf{\bibinfo{volume}{88}},
  \bibinfo{pages}{111501} (\bibinfo{year}{2013}{\natexlab{b}}),
  \eprint{1309.7644}.

\bibitem[{\citenamefont{Grabovsky}(2013)}]{grabovsky:2013mba}
\bibinfo{author}{\bibfnamefont{A.~V.} \bibnamefont{Grabovsky}},
  \bibinfo{journal}{JHEP} \textbf{\bibinfo{volume}{09}}, \bibinfo{pages}{141}
  (\bibinfo{year}{2013}), \eprint{1307.5414}.

\bibitem[{\citenamefont{Caron-Huot}(2015)}]{caron-huot:2013fea}
\bibinfo{author}{\bibfnamefont{S.}~\bibnamefont{Caron-Huot}},
  \bibinfo{journal}{JHEP} \textbf{\bibinfo{volume}{05}}, \bibinfo{pages}{093}
  (\bibinfo{year}{2015}), \eprint{1309.6521}.

\bibitem[{\citenamefont{Kovner et~al.}(2014)\citenamefont{Kovner, Lublinsky,
  and Mulian}}]{kovner:2013ona}
\bibinfo{author}{\bibfnamefont{A.}~\bibnamefont{Kovner}},
  \bibinfo{author}{\bibfnamefont{M.}~\bibnamefont{Lublinsky}},
  \bibnamefont{and} \bibinfo{author}{\bibfnamefont{Y.}~\bibnamefont{Mulian}},
  \bibinfo{journal}{Phys. Rev. D} \textbf{\bibinfo{volume}{89}},
  \bibinfo{pages}{061704} (\bibinfo{year}{2014}), \eprint{1310.0378}.

\bibitem[{\citenamefont{Lublinsky and Mulian}(2017)}]{lublinsky:2016meo}
\bibinfo{author}{\bibfnamefont{M.}~\bibnamefont{Lublinsky}} \bibnamefont{and}
  \bibinfo{author}{\bibfnamefont{Y.}~\bibnamefont{Mulian}},
  \bibinfo{journal}{JHEP} \textbf{\bibinfo{volume}{05}}, \bibinfo{pages}{097}
  (\bibinfo{year}{2017}), \eprint{1610.03453}.

\bibitem[{\citenamefont{Caucal et~al.}(2022)\citenamefont{Caucal, Salazar,
  Schenke, and Venugopalan}}]{Caucal:2022ulg}
\bibinfo{author}{\bibfnamefont{P.}~\bibnamefont{Caucal}},
  \bibinfo{author}{\bibfnamefont{F.}~\bibnamefont{Salazar}},
  \bibinfo{author}{\bibfnamefont{B.}~\bibnamefont{Schenke}}, \bibnamefont{and}
  \bibinfo{author}{\bibfnamefont{R.}~\bibnamefont{Venugopalan}},
  \bibinfo{journal}{JHEP} \textbf{\bibinfo{volume}{11}}, \bibinfo{pages}{169}
  (\bibinfo{year}{2022}), \eprint{2208.13872}.

\bibitem[{\citenamefont{Caucal et~al.}(2023{\natexlab{a}})\citenamefont{Caucal,
  Salazar, Schenke, Stebel, and Venugopalan}}]{Caucal:2023nci}
\bibinfo{author}{\bibfnamefont{P.}~\bibnamefont{Caucal}},
  \bibinfo{author}{\bibfnamefont{F.}~\bibnamefont{Salazar}},
  \bibinfo{author}{\bibfnamefont{B.}~\bibnamefont{Schenke}},
  \bibinfo{author}{\bibfnamefont{T.}~\bibnamefont{Stebel}}, \bibnamefont{and}
  \bibinfo{author}{\bibfnamefont{R.}~\bibnamefont{Venugopalan}},
  \bibinfo{journal}{JHEP} \textbf{\bibinfo{volume}{08}}, \bibinfo{pages}{062}
  (\bibinfo{year}{2023}{\natexlab{a}}), \eprint{2304.03304}.

\bibitem[{\citenamefont{Caucal et~al.}(2023{\natexlab{b}})\citenamefont{Caucal,
  Salazar, Schenke, Stebel, and Venugopalan}}]{Caucal:2023fsf}
\bibinfo{author}{\bibfnamefont{P.}~\bibnamefont{Caucal}},
  \bibinfo{author}{\bibfnamefont{F.}~\bibnamefont{Salazar}},
  \bibinfo{author}{\bibfnamefont{B.}~\bibnamefont{Schenke}},
  \bibinfo{author}{\bibfnamefont{T.}~\bibnamefont{Stebel}}, \bibnamefont{and}
  \bibinfo{author}{\bibfnamefont{R.}~\bibnamefont{Venugopalan}}
  (\bibinfo{year}{2023}{\natexlab{b}}), \eprint{2308.00022}.

\bibitem[{\citenamefont{Caron-Huot and Herranen}(2018)}]{caron-huot:2016tzz}
\bibinfo{author}{\bibfnamefont{S.}~\bibnamefont{Caron-Huot}} \bibnamefont{and}
  \bibinfo{author}{\bibfnamefont{M.}~\bibnamefont{Herranen}},
  \bibinfo{journal}{JHEP} \textbf{\bibinfo{volume}{02}}, \bibinfo{pages}{058}
  (\bibinfo{year}{2018}), \eprint{1604.07417}.

\bibitem[{\citenamefont{Fucilla
  et~al.}(2023{\natexlab{a}})\citenamefont{Fucilla, Grabovsky, Li, Szymanowski,
  and Wallon}}]{Fucilla:2023mkl}
\bibinfo{author}{\bibfnamefont{M.}~\bibnamefont{Fucilla}},
  \bibinfo{author}{\bibfnamefont{A.}~\bibnamefont{Grabovsky}},
  \bibinfo{author}{\bibfnamefont{E.}~\bibnamefont{Li}},
  \bibinfo{author}{\bibfnamefont{L.}~\bibnamefont{Szymanowski}},
  \bibnamefont{and} \bibinfo{author}{\bibfnamefont{S.}~\bibnamefont{Wallon}}
  (\bibinfo{year}{2023}{\natexlab{a}}), \eprint{2310.11066}.

\bibitem[{\citenamefont{Fucilla
  et~al.}(2023{\natexlab{b}})\citenamefont{Fucilla, Grabovsky, Li, Szymanowski,
  and Wallon}}]{Fucilla:2022wcg}
\bibinfo{author}{\bibfnamefont{M.}~\bibnamefont{Fucilla}},
  \bibinfo{author}{\bibfnamefont{A.~V.} \bibnamefont{Grabovsky}},
  \bibinfo{author}{\bibfnamefont{E.}~\bibnamefont{Li}},
  \bibinfo{author}{\bibfnamefont{L.}~\bibnamefont{Szymanowski}},
  \bibnamefont{and} \bibinfo{author}{\bibfnamefont{S.}~\bibnamefont{Wallon}},
  \bibinfo{journal}{JHEP} \textbf{\bibinfo{volume}{03}}, \bibinfo{pages}{159}
  (\bibinfo{year}{2023}{\natexlab{b}}), \eprint{2211.05774}.

\bibitem[{\citenamefont{Boussarie et~al.}(2018)\citenamefont{Boussarie,
  Grabovsky, Ivanov, Szymanowski, and Wallon}}]{boussarie:2017dmx}
\bibinfo{author}{\bibfnamefont{R.}~\bibnamefont{Boussarie}},
  \bibinfo{author}{\bibfnamefont{A.~V.} \bibnamefont{Grabovsky}},
  \bibinfo{author}{\bibfnamefont{D.~Y.} \bibnamefont{Ivanov}},
  \bibinfo{author}{\bibfnamefont{L.}~\bibnamefont{Szymanowski}},
  \bibnamefont{and} \bibinfo{author}{\bibfnamefont{S.}~\bibnamefont{Wallon}},
  \bibinfo{journal}{PoS} \textbf{\bibinfo{volume}{DIS2017}},
  \bibinfo{pages}{062} (\bibinfo{year}{2018}), \eprint{1709.04422}.

\bibitem[{\citenamefont{Beuf et~al.}(2022{\natexlab{a}})\citenamefont{Beuf,
  Lappi, and Paatelainen}}]{beuf:2022ndu}
\bibinfo{author}{\bibfnamefont{G.}~\bibnamefont{Beuf}},
  \bibinfo{author}{\bibfnamefont{T.}~\bibnamefont{Lappi}}, \bibnamefont{and}
  \bibinfo{author}{\bibfnamefont{R.}~\bibnamefont{Paatelainen}},
  \bibinfo{journal}{Phys. Rev. D} \textbf{\bibinfo{volume}{106}},
  \bibinfo{pages}{034013} (\bibinfo{year}{2022}{\natexlab{a}}),
  \eprint{2204.02486}.

\bibitem[{\citenamefont{Beuf et~al.}(2022{\natexlab{b}})\citenamefont{Beuf,
  Lappi, and Paatelainen}}]{beuf:2021srj}
\bibinfo{author}{\bibfnamefont{G.}~\bibnamefont{Beuf}},
  \bibinfo{author}{\bibfnamefont{T.}~\bibnamefont{Lappi}}, \bibnamefont{and}
  \bibinfo{author}{\bibfnamefont{R.}~\bibnamefont{Paatelainen}},
  \bibinfo{journal}{Phys. Rev. Lett.} \textbf{\bibinfo{volume}{129}},
  \bibinfo{pages}{072001} (\bibinfo{year}{2022}{\natexlab{b}}),
  \eprint{2112.03158}.

\bibitem[{\citenamefont{Beuf et~al.}(2021)\citenamefont{Beuf, Lappi, and
  Paatelainen}}]{beuf:2021qqa}
\bibinfo{author}{\bibfnamefont{G.}~\bibnamefont{Beuf}},
  \bibinfo{author}{\bibfnamefont{T.}~\bibnamefont{Lappi}}, \bibnamefont{and}
  \bibinfo{author}{\bibfnamefont{R.}~\bibnamefont{Paatelainen}},
  \bibinfo{journal}{Phys. Rev. D} \textbf{\bibinfo{volume}{104}},
  \bibinfo{pages}{056032} (\bibinfo{year}{2021}), \eprint{2103.14549}.

\bibitem[{\citenamefont{M\"antysaari and Penttala}(2021)}]{mantysaari:2021ryb}
\bibinfo{author}{\bibfnamefont{H.}~\bibnamefont{M\"antysaari}}
  \bibnamefont{and} \bibinfo{author}{\bibfnamefont{J.}~\bibnamefont{Penttala}},
  \bibinfo{journal}{Phys. Lett. B} \textbf{\bibinfo{volume}{823}},
  \bibinfo{pages}{136723} (\bibinfo{year}{2021}), \eprint{2104.02349}.

\bibitem[{\citenamefont{M\"antysaari and
  Penttala}(2022{\natexlab{a}})}]{mantysaari:2022bsp}
\bibinfo{author}{\bibfnamefont{H.}~\bibnamefont{M\"antysaari}}
  \bibnamefont{and} \bibinfo{author}{\bibfnamefont{J.}~\bibnamefont{Penttala}},
  \bibinfo{journal}{Phys. Rev. D} \textbf{\bibinfo{volume}{105}},
  \bibinfo{pages}{114038} (\bibinfo{year}{2022}{\natexlab{a}}),
  \eprint{2203.16911}.

\bibitem[{\citenamefont{M\"antysaari and
  Penttala}(2022{\natexlab{b}})}]{mantysaari:2022kdm}
\bibinfo{author}{\bibfnamefont{H.}~\bibnamefont{M\"antysaari}}
  \bibnamefont{and} \bibinfo{author}{\bibfnamefont{J.}~\bibnamefont{Penttala}},
  \bibinfo{journal}{JHEP} \textbf{\bibinfo{volume}{08}}, \bibinfo{pages}{247}
  (\bibinfo{year}{2022}{\natexlab{b}}), \eprint{2204.14031}.

\bibitem[{\citenamefont{Lappi et~al.}(2022)\citenamefont{Lappi, M\"antysaari,
  and Penttala}}]{lappi:2021oag}
\bibinfo{author}{\bibfnamefont{T.}~\bibnamefont{Lappi}},
  \bibinfo{author}{\bibfnamefont{H.}~\bibnamefont{M\"antysaari}},
  \bibnamefont{and} \bibinfo{author}{\bibfnamefont{J.}~\bibnamefont{Penttala}},
  \bibinfo{journal}{SciPost Phys. Proc.} \textbf{\bibinfo{volume}{8}},
  \bibinfo{pages}{133} (\bibinfo{year}{2022}), \eprint{2106.12825}.

\bibitem[{\citenamefont{Ayala et~al.}(2016)\citenamefont{Ayala, Hentschinski,
  Jalilian-Marian, and Tejeda-Yeomans}}]{Ayala:2016lhd}
\bibinfo{author}{\bibfnamefont{A.}~\bibnamefont{Ayala}},
  \bibinfo{author}{\bibfnamefont{M.}~\bibnamefont{Hentschinski}},
  \bibinfo{author}{\bibfnamefont{J.}~\bibnamefont{Jalilian-Marian}},
  \bibnamefont{and} \bibinfo{author}{\bibfnamefont{M.~E.}
  \bibnamefont{Tejeda-Yeomans}}, \bibinfo{journal}{Phys. Lett. B}
  \textbf{\bibinfo{volume}{761}}, \bibinfo{pages}{229} (\bibinfo{year}{2016}),
  \eprint{1604.08526}.

\bibitem[{\citenamefont{Ayala et~al.}(2017)\citenamefont{Ayala, Hentschinski,
  Jalilian-Marian, and Tejeda-Yeomans}}]{Ayala:2017rmh}
\bibinfo{author}{\bibfnamefont{A.}~\bibnamefont{Ayala}},
  \bibinfo{author}{\bibfnamefont{M.}~\bibnamefont{Hentschinski}},
  \bibinfo{author}{\bibfnamefont{J.}~\bibnamefont{Jalilian-Marian}},
  \bibnamefont{and} \bibinfo{author}{\bibfnamefont{M.~E.}
  \bibnamefont{Tejeda-Yeomans}}, \bibinfo{journal}{Nucl. Phys. B}
  \textbf{\bibinfo{volume}{920}}, \bibinfo{pages}{232} (\bibinfo{year}{2017}),
  \eprint{1701.07143}.

\bibitem[{\citenamefont{Ayala et~al.}(2014)\citenamefont{Ayala, Cazaroto,
  Hern\'andez, Jalilian-Marian, and Tejeda-Yeomans}}]{Ayala:2014nza}
\bibinfo{author}{\bibfnamefont{A.}~\bibnamefont{Ayala}},
  \bibinfo{author}{\bibfnamefont{E.~R.} \bibnamefont{Cazaroto}},
  \bibinfo{author}{\bibfnamefont{L.~A.} \bibnamefont{Hern\'andez}},
  \bibinfo{author}{\bibfnamefont{J.}~\bibnamefont{Jalilian-Marian}},
  \bibnamefont{and} \bibinfo{author}{\bibfnamefont{M.~E.}
  \bibnamefont{Tejeda-Yeomans}}, \bibinfo{journal}{Phys. Rev. D}
  \textbf{\bibinfo{volume}{90}}, \bibinfo{pages}{074037}
  (\bibinfo{year}{2014}), \eprint{1408.3080}.

\bibitem[{\citenamefont{Iancu and Mulian}(2021)}]{Iancu:2020mos}
\bibinfo{author}{\bibfnamefont{E.}~\bibnamefont{Iancu}} \bibnamefont{and}
  \bibinfo{author}{\bibfnamefont{Y.}~\bibnamefont{Mulian}},
  \bibinfo{journal}{JHEP} \textbf{\bibinfo{volume}{03}}, \bibinfo{pages}{005}
  (\bibinfo{year}{2021}), \eprint{2009.11930}.

\bibitem[{\citenamefont{Roy and Venugopalan}(2020)}]{roy:2019hwr}
\bibinfo{author}{\bibfnamefont{K.}~\bibnamefont{Roy}} \bibnamefont{and}
  \bibinfo{author}{\bibfnamefont{R.}~\bibnamefont{Venugopalan}},
  \bibinfo{journal}{Phys. Rev. D} \textbf{\bibinfo{volume}{101}},
  \bibinfo{pages}{034028} (\bibinfo{year}{2020}), \eprint{1911.04530}.

\bibitem[{\citenamefont{Hatta et~al.}(2022)\citenamefont{Hatta, Xiao, and
  Yuan}}]{hatta:2022lzj}
\bibinfo{author}{\bibfnamefont{Y.}~\bibnamefont{Hatta}},
  \bibinfo{author}{\bibfnamefont{B.-W.} \bibnamefont{Xiao}}, \bibnamefont{and}
  \bibinfo{author}{\bibfnamefont{F.}~\bibnamefont{Yuan}}
  (\bibinfo{year}{2022}), \eprint{2205.08060}.

\bibitem[{\citenamefont{Iancu et~al.}(2022)\citenamefont{Iancu, Mueller, and
  Triantafyllopoulos}}]{Iancu:2021rup}
\bibinfo{author}{\bibfnamefont{E.}~\bibnamefont{Iancu}},
  \bibinfo{author}{\bibfnamefont{A.~H.} \bibnamefont{Mueller}},
  \bibnamefont{and} \bibinfo{author}{\bibfnamefont{D.~N.}
  \bibnamefont{Triantafyllopoulos}}, \bibinfo{journal}{Phys. Rev. Lett.}
  \textbf{\bibinfo{volume}{128}}, \bibinfo{pages}{202001}
  (\bibinfo{year}{2022}), \eprint{2112.06353}.

\bibitem[{\citenamefont{Iancu et~al.}(2021)\citenamefont{Iancu, Mueller,
  Triantafyllopoulos, and Wei}}]{Iancu:2020jch}
\bibinfo{author}{\bibfnamefont{E.}~\bibnamefont{Iancu}},
  \bibinfo{author}{\bibfnamefont{A.~H.} \bibnamefont{Mueller}},
  \bibinfo{author}{\bibfnamefont{D.~N.} \bibnamefont{Triantafyllopoulos}},
  \bibnamefont{and} \bibinfo{author}{\bibfnamefont{S.~Y.} \bibnamefont{Wei}},
  \bibinfo{journal}{JHEP} \textbf{\bibinfo{volume}{07}}, \bibinfo{pages}{196}
  (\bibinfo{year}{2021}), \eprint{2012.08562}.

\bibitem[{\citenamefont{Taels et~al.}(2022)\citenamefont{Taels, Altinoluk,
  Beuf, and Marquet}}]{Taels:2022tza}
\bibinfo{author}{\bibfnamefont{P.}~\bibnamefont{Taels}},
  \bibinfo{author}{\bibfnamefont{T.}~\bibnamefont{Altinoluk}},
  \bibinfo{author}{\bibfnamefont{G.}~\bibnamefont{Beuf}}, \bibnamefont{and}
  \bibinfo{author}{\bibfnamefont{C.}~\bibnamefont{Marquet}}
  (\bibinfo{year}{2022}), \eprint{2204.11650}.

\bibitem[{\citenamefont{Caucal et~al.}(2021)\citenamefont{Caucal, Salazar, and
  Venugopalan}}]{Caucal:2021ent}
\bibinfo{author}{\bibfnamefont{P.}~\bibnamefont{Caucal}},
  \bibinfo{author}{\bibfnamefont{F.}~\bibnamefont{Salazar}}, \bibnamefont{and}
  \bibinfo{author}{\bibfnamefont{R.}~\bibnamefont{Venugopalan}},
  \bibinfo{journal}{JHEP} \textbf{\bibinfo{volume}{11}}, \bibinfo{pages}{222}
  (\bibinfo{year}{2021}), \eprint{2108.06347}.

\bibitem[{\citenamefont{Bergabo and
  Jalilian-Marian}(2022{\natexlab{a}})}]{Bergabo:2021woe}
\bibinfo{author}{\bibfnamefont{F.}~\bibnamefont{Bergabo}} \bibnamefont{and}
  \bibinfo{author}{\bibfnamefont{J.}~\bibnamefont{Jalilian-Marian}},
  \bibinfo{journal}{Nucl. Phys. A} \textbf{\bibinfo{volume}{1018}},
  \bibinfo{pages}{122358} (\bibinfo{year}{2022}{\natexlab{a}}),
  \eprint{2108.10428}.

\bibitem[{\citenamefont{Bergabo and
  Jalilian-Marian}(2022{\natexlab{b}})}]{Bergabo:2022tcu}
\bibinfo{author}{\bibfnamefont{F.}~\bibnamefont{Bergabo}} \bibnamefont{and}
  \bibinfo{author}{\bibfnamefont{J.}~\bibnamefont{Jalilian-Marian}},
  \bibinfo{journal}{Phys. Rev. D} \textbf{\bibinfo{volume}{106}},
  \bibinfo{pages}{054035} (\bibinfo{year}{2022}{\natexlab{b}}),
  \eprint{2207.03606}.

\bibitem[{\citenamefont{Kovner and Wiedemann}(2001)}]{Kovner:2001vi}
\bibinfo{author}{\bibfnamefont{A.}~\bibnamefont{Kovner}} \bibnamefont{and}
  \bibinfo{author}{\bibfnamefont{U.~A.} \bibnamefont{Wiedemann}},
  \bibinfo{journal}{Phys. Rev. D} \textbf{\bibinfo{volume}{64}},
  \bibinfo{pages}{114002} (\bibinfo{year}{2001}), \eprint{hep-ph/0106240}.

\bibitem[{\citenamefont{Kovchegov
  et~al.}(2017{\natexlab{a}})\citenamefont{Kovchegov, Pitonyak, and
  Sievert}}]{Kovchegov:2017lsr}
\bibinfo{author}{\bibfnamefont{Y.~V.} \bibnamefont{Kovchegov}},
  \bibinfo{author}{\bibfnamefont{D.}~\bibnamefont{Pitonyak}}, \bibnamefont{and}
  \bibinfo{author}{\bibfnamefont{M.~D.} \bibnamefont{Sievert}},
  \bibinfo{journal}{JHEP} \textbf{\bibinfo{volume}{10}}, \bibinfo{pages}{198}
  (\bibinfo{year}{2017}{\natexlab{a}}), \eprint{1706.04236}.

\bibitem[{\citenamefont{Cougoulic and Kovchegov}(2019)}]{Cougoulic:2019aja}
\bibinfo{author}{\bibfnamefont{F.}~\bibnamefont{Cougoulic}} \bibnamefont{and}
  \bibinfo{author}{\bibfnamefont{Y.~V.} \bibnamefont{Kovchegov}},
  \bibinfo{journal}{Phys. Rev. D} \textbf{\bibinfo{volume}{100}},
  \bibinfo{pages}{114020} (\bibinfo{year}{2019}), \eprint{1910.04268}.

\bibitem[{\citenamefont{Kovchegov and Sievert}(2019)}]{Kovchegov:2018znm}
\bibinfo{author}{\bibfnamefont{Y.~V.} \bibnamefont{Kovchegov}}
  \bibnamefont{and} \bibinfo{author}{\bibfnamefont{M.~D.}
  \bibnamefont{Sievert}}, \bibinfo{journal}{Phys. Rev. D}
  \textbf{\bibinfo{volume}{99}}, \bibinfo{pages}{054032}
  (\bibinfo{year}{2019}), \eprint{1808.09010}.

\bibitem[{\citenamefont{Kovchegov
  et~al.}(2017{\natexlab{b}})\citenamefont{Kovchegov, Pitonyak, and
  Sievert}}]{Kovchegov:2017jxc}
\bibinfo{author}{\bibfnamefont{Y.~V.} \bibnamefont{Kovchegov}},
  \bibinfo{author}{\bibfnamefont{D.}~\bibnamefont{Pitonyak}}, \bibnamefont{and}
  \bibinfo{author}{\bibfnamefont{M.~D.} \bibnamefont{Sievert}},
  \bibinfo{journal}{Phys. Lett. B} \textbf{\bibinfo{volume}{772}},
  \bibinfo{pages}{136} (\bibinfo{year}{2017}{\natexlab{b}}),
  \eprint{1703.05809}.

\bibitem[{\citenamefont{Kovchegov
  et~al.}(2017{\natexlab{c}})\citenamefont{Kovchegov, Pitonyak, and
  Sievert}}]{Kovchegov:2016zex}
\bibinfo{author}{\bibfnamefont{Y.~V.} \bibnamefont{Kovchegov}},
  \bibinfo{author}{\bibfnamefont{D.}~\bibnamefont{Pitonyak}}, \bibnamefont{and}
  \bibinfo{author}{\bibfnamefont{M.~D.} \bibnamefont{Sievert}},
  \bibinfo{journal}{Phys. Rev. D} \textbf{\bibinfo{volume}{95}},
  \bibinfo{pages}{014033} (\bibinfo{year}{2017}{\natexlab{c}}),
  \eprint{1610.06197}.

\bibitem[{\citenamefont{Kovchegov
  et~al.}(2017{\natexlab{d}})\citenamefont{Kovchegov, Pitonyak, and
  Sievert}}]{Kovchegov:2016weo}
\bibinfo{author}{\bibfnamefont{Y.~V.} \bibnamefont{Kovchegov}},
  \bibinfo{author}{\bibfnamefont{D.}~\bibnamefont{Pitonyak}}, \bibnamefont{and}
  \bibinfo{author}{\bibfnamefont{M.~D.} \bibnamefont{Sievert}},
  \bibinfo{journal}{Phys. Rev. Lett.} \textbf{\bibinfo{volume}{118}},
  \bibinfo{pages}{052001} (\bibinfo{year}{2017}{\natexlab{d}}),
  \eprint{1610.06188}.

\bibitem[{\citenamefont{Kovchegov et~al.}(2016)\citenamefont{Kovchegov,
  Pitonyak, and Sievert}}]{Kovchegov:2015pbl}
\bibinfo{author}{\bibfnamefont{Y.~V.} \bibnamefont{Kovchegov}},
  \bibinfo{author}{\bibfnamefont{D.}~\bibnamefont{Pitonyak}}, \bibnamefont{and}
  \bibinfo{author}{\bibfnamefont{M.~D.} \bibnamefont{Sievert}},
  \bibinfo{journal}{JHEP} \textbf{\bibinfo{volume}{01}}, \bibinfo{pages}{072}
  (\bibinfo{year}{2016}), \bibinfo{note}{[Erratum: JHEP 10, 148 (2016)]},
  \eprint{1511.06737}.

\bibitem[{\citenamefont{Agostini
  et~al.}(2019{\natexlab{a}})\citenamefont{Agostini, Altinoluk, and
  Armesto}}]{Agostini:2019hkj}
\bibinfo{author}{\bibfnamefont{P.}~\bibnamefont{Agostini}},
  \bibinfo{author}{\bibfnamefont{T.}~\bibnamefont{Altinoluk}},
  \bibnamefont{and} \bibinfo{author}{\bibfnamefont{N.}~\bibnamefont{Armesto}},
  \bibinfo{journal}{Eur. Phys. J. C} \textbf{\bibinfo{volume}{79}},
  \bibinfo{pages}{790} (\bibinfo{year}{2019}{\natexlab{a}}),
  \eprint{1907.03668}.

\bibitem[{\citenamefont{Agostini
  et~al.}(2019{\natexlab{b}})\citenamefont{Agostini, Altinoluk, and
  Armesto}}]{Agostini:2019avp}
\bibinfo{author}{\bibfnamefont{P.}~\bibnamefont{Agostini}},
  \bibinfo{author}{\bibfnamefont{T.}~\bibnamefont{Altinoluk}},
  \bibnamefont{and} \bibinfo{author}{\bibfnamefont{N.}~\bibnamefont{Armesto}},
  \bibinfo{journal}{Eur. Phys. J. C} \textbf{\bibinfo{volume}{79}},
  \bibinfo{pages}{600} (\bibinfo{year}{2019}{\natexlab{b}}),
  \eprint{1902.04483}.

\bibitem[{\citenamefont{Altinoluk and Dumitru}(2016)}]{Altinoluk:2015xuy}
\bibinfo{author}{\bibfnamefont{T.}~\bibnamefont{Altinoluk}} \bibnamefont{and}
  \bibinfo{author}{\bibfnamefont{A.}~\bibnamefont{Dumitru}},
  \bibinfo{journal}{Phys. Rev. D} \textbf{\bibinfo{volume}{94}},
  \bibinfo{pages}{074032} (\bibinfo{year}{2016}), \eprint{1512.00279}.

\bibitem[{\citenamefont{Altinoluk
  et~al.}(2016{\natexlab{b}})\citenamefont{Altinoluk, Armesto, Beuf, and
  Moscoso}}]{Altinoluk:2015gia}
\bibinfo{author}{\bibfnamefont{T.}~\bibnamefont{Altinoluk}},
  \bibinfo{author}{\bibfnamefont{N.}~\bibnamefont{Armesto}},
  \bibinfo{author}{\bibfnamefont{G.}~\bibnamefont{Beuf}}, \bibnamefont{and}
  \bibinfo{author}{\bibfnamefont{A.}~\bibnamefont{Moscoso}},
  \bibinfo{journal}{JHEP} \textbf{\bibinfo{volume}{01}}, \bibinfo{pages}{114}
  (\bibinfo{year}{2016}{\natexlab{b}}), \eprint{1505.01400}.

\bibitem[{\citenamefont{Altinoluk et~al.}(2014)\citenamefont{Altinoluk,
  Armesto, Beuf, Mart\'\i{}nez, and Salgado}}]{Altinoluk:2014oxa}
\bibinfo{author}{\bibfnamefont{T.}~\bibnamefont{Altinoluk}},
  \bibinfo{author}{\bibfnamefont{N.}~\bibnamefont{Armesto}},
  \bibinfo{author}{\bibfnamefont{G.}~\bibnamefont{Beuf}},
  \bibinfo{author}{\bibfnamefont{M.}~\bibnamefont{Mart\'\i{}nez}},
  \bibnamefont{and} \bibinfo{author}{\bibfnamefont{C.~A.}
  \bibnamefont{Salgado}}, \bibinfo{journal}{JHEP}
  \textbf{\bibinfo{volume}{07}}, \bibinfo{pages}{068} (\bibinfo{year}{2014}),
  \eprint{1404.2219}.

\bibitem[{\citenamefont{Jalilian-Marian}(2021)}]{jalilian-marian:2021lhe}
\bibinfo{author}{\bibfnamefont{J.}~\bibnamefont{Jalilian-Marian}},
  \bibinfo{journal}{Nucl. Phys. A} \textbf{\bibinfo{volume}{1005}},
  \bibinfo{pages}{121943} (\bibinfo{year}{2021}).

\bibitem[{\citenamefont{Jalilian-Marian}(2020)}]{Jalilian-Marian:2019kaf}
\bibinfo{author}{\bibfnamefont{J.}~\bibnamefont{Jalilian-Marian}},
  \bibinfo{journal}{Phys. Rev. D} \textbf{\bibinfo{volume}{102}},
  \bibinfo{pages}{014008} (\bibinfo{year}{2020}), \eprint{1912.08878}.

\bibitem[{\citenamefont{Jalilian-Marian}(2019)}]{Jalilian-Marian:2018iui}
\bibinfo{author}{\bibfnamefont{J.}~\bibnamefont{Jalilian-Marian}},
  \bibinfo{journal}{Phys. Rev. D} \textbf{\bibinfo{volume}{99}},
  \bibinfo{pages}{014043} (\bibinfo{year}{2019}), \eprint{1809.04625}.

\bibitem[{\citenamefont{Jalilian-Marian}(2017)}]{Jalilian-Marian:2017ttv}
\bibinfo{author}{\bibfnamefont{J.}~\bibnamefont{Jalilian-Marian}},
  \bibinfo{journal}{Phys. Rev. D} \textbf{\bibinfo{volume}{96}},
  \bibinfo{pages}{074020} (\bibinfo{year}{2017}), \eprint{1708.07533}.

\bibitem[{\citenamefont{Hentschinski et~al.}(2018)\citenamefont{Hentschinski,
  Kusina, Kutak, and Serino}}]{Hentschinski:2017ayz}
\bibinfo{author}{\bibfnamefont{M.}~\bibnamefont{Hentschinski}},
  \bibinfo{author}{\bibfnamefont{A.}~\bibnamefont{Kusina}},
  \bibinfo{author}{\bibfnamefont{K.}~\bibnamefont{Kutak}}, \bibnamefont{and}
  \bibinfo{author}{\bibfnamefont{M.}~\bibnamefont{Serino}},
  \bibinfo{journal}{Eur. Phys. J. C} \textbf{\bibinfo{volume}{78}},
  \bibinfo{pages}{174} (\bibinfo{year}{2018}), \eprint{1711.04587}.

\bibitem[{\citenamefont{Hentschinski et~al.}(2016)\citenamefont{Hentschinski,
  Kusina, and Kutak}}]{Hentschinski:2016wya}
\bibinfo{author}{\bibfnamefont{M.}~\bibnamefont{Hentschinski}},
  \bibinfo{author}{\bibfnamefont{A.}~\bibnamefont{Kusina}}, \bibnamefont{and}
  \bibinfo{author}{\bibfnamefont{K.}~\bibnamefont{Kutak}},
  \bibinfo{journal}{Phys. Rev. D} \textbf{\bibinfo{volume}{94}},
  \bibinfo{pages}{114013} (\bibinfo{year}{2016}), \eprint{1607.01507}.

\bibitem[{\citenamefont{Gituliar et~al.}(2016)\citenamefont{Gituliar,
  Hentschinski, and Kutak}}]{Gituliar:2015agu}
\bibinfo{author}{\bibfnamefont{O.}~\bibnamefont{Gituliar}},
  \bibinfo{author}{\bibfnamefont{M.}~\bibnamefont{Hentschinski}},
  \bibnamefont{and} \bibinfo{author}{\bibfnamefont{K.}~\bibnamefont{Kutak}},
  \bibinfo{journal}{JHEP} \textbf{\bibinfo{volume}{01}}, \bibinfo{pages}{181}
  (\bibinfo{year}{2016}), \eprint{1511.08439}.

\bibitem[{\citenamefont{Balitsky and Tarasov}(2016)}]{Balitsky:2016dgz}
\bibinfo{author}{\bibfnamefont{I.}~\bibnamefont{Balitsky}} \bibnamefont{and}
  \bibinfo{author}{\bibfnamefont{A.}~\bibnamefont{Tarasov}},
  \bibinfo{journal}{JHEP} \textbf{\bibinfo{volume}{06}}, \bibinfo{pages}{164}
  (\bibinfo{year}{2016}), \eprint{1603.06548}.

\bibitem[{\citenamefont{Balitsky and Tarasov}(2015)}]{Balitsky:2015qba}
\bibinfo{author}{\bibfnamefont{I.}~\bibnamefont{Balitsky}} \bibnamefont{and}
  \bibinfo{author}{\bibfnamefont{A.}~\bibnamefont{Tarasov}},
  \bibinfo{journal}{JHEP} \textbf{\bibinfo{volume}{10}}, \bibinfo{pages}{017}
  (\bibinfo{year}{2015}), \eprint{1505.02151}.

\bibitem[{\citenamefont{Mukherjee et~al.}(2023)\citenamefont{Mukherjee, Skokov,
  Tarasov, and Tiwari}}]{Mukherjee:2023snp}
\bibinfo{author}{\bibfnamefont{S.}~\bibnamefont{Mukherjee}},
  \bibinfo{author}{\bibfnamefont{V.~V.} \bibnamefont{Skokov}},
  \bibinfo{author}{\bibfnamefont{A.}~\bibnamefont{Tarasov}}, \bibnamefont{and}
  \bibinfo{author}{\bibfnamefont{S.}~\bibnamefont{Tiwari}}
  (\bibinfo{year}{2023}), \eprint{2311.16402}.

\bibitem[{\citenamefont{Fu et~al.}(2023)\citenamefont{Fu, Kang, Salazar, Wang,
  and Xing}}]{Fu:2023jqv}
\bibinfo{author}{\bibfnamefont{Y.}~\bibnamefont{Fu}},
  \bibinfo{author}{\bibfnamefont{Z.-B.} \bibnamefont{Kang}},
  \bibinfo{author}{\bibfnamefont{F.}~\bibnamefont{Salazar}},
  \bibinfo{author}{\bibfnamefont{X.-N.} \bibnamefont{Wang}}, \bibnamefont{and}
  \bibinfo{author}{\bibfnamefont{H.}~\bibnamefont{Xing}}
  (\bibinfo{year}{2023}), \eprint{2310.12847}.

\bibitem[{\citenamefont{Boussarie and Mehtar-Tani}(2022)}]{Boussarie:2021wkn}
\bibinfo{author}{\bibfnamefont{R.}~\bibnamefont{Boussarie}} \bibnamefont{and}
  \bibinfo{author}{\bibfnamefont{Y.}~\bibnamefont{Mehtar-Tani}},
  \bibinfo{journal}{JHEP} \textbf{\bibinfo{volume}{07}}, \bibinfo{pages}{080}
  (\bibinfo{year}{2022}), \eprint{2112.01412}.

\bibitem[{\citenamefont{Boussarie and Mehtar-Tani}(2023)}]{Boussarie:2023xun}
\bibinfo{author}{\bibfnamefont{R.}~\bibnamefont{Boussarie}} \bibnamefont{and}
  \bibinfo{author}{\bibfnamefont{Y.}~\bibnamefont{Mehtar-Tani}}
  (\bibinfo{year}{2023}), \eprint{2309.16576}.

\bibitem[{\citenamefont{Bergabo and
  Jalilian-Marian}(2023{\natexlab{a}})}]{Bergabo:2022zhe}
\bibinfo{author}{\bibfnamefont{F.}~\bibnamefont{Bergabo}} \bibnamefont{and}
  \bibinfo{author}{\bibfnamefont{J.}~\bibnamefont{Jalilian-Marian}},
  \bibinfo{journal}{JHEP} \textbf{\bibinfo{volume}{01}}, \bibinfo{pages}{095}
  (\bibinfo{year}{2023}{\natexlab{a}}), \eprint{2210.03208}.

\bibitem[{\citenamefont{Caucal et~al.}(2024)\citenamefont{Caucal, Ferrand, and
  Salazar}}]{Caucal:2024cdq}
\bibinfo{author}{\bibfnamefont{P.}~\bibnamefont{Caucal}},
  \bibinfo{author}{\bibfnamefont{E.}~\bibnamefont{Ferrand}}, \bibnamefont{and}
  \bibinfo{author}{\bibfnamefont{F.}~\bibnamefont{Salazar}}
  (\bibinfo{year}{2024}), \eprint{2401.01934}.

\bibitem[{\citenamefont{Bergabo and
  Jalilian-Marian}(2023{\natexlab{b}})}]{Bergabo:2023wed}
\bibinfo{author}{\bibfnamefont{F.}~\bibnamefont{Bergabo}} \bibnamefont{and}
  \bibinfo{author}{\bibfnamefont{J.}~\bibnamefont{Jalilian-Marian}},
  \bibinfo{journal}{Phys. Rev. D} \textbf{\bibinfo{volume}{107}},
  \bibinfo{pages}{054036} (\bibinfo{year}{2023}{\natexlab{b}}),
  \eprint{2301.03117}.

\end{thebibliography}
\bibliographystyle{apsrev}

\end{document}